\documentclass[aps,showpacs,preprintnumbers,amsmath, amssymb]{revtex4}

\usepackage{float}
\usepackage{graphics,epsfig}
\usepackage{graphicx}
\usepackage{dcolumn}
\usepackage{bm}

\begin{document}
\baselineskip=0.8 cm
\title{{\bf Excited states of holographic superconductors with hyperscaling violation}}

\author{Shuhang Zhang$^{1}$, Zixu Zhao$^{1}$\footnote{Corresponding author: zhao$_{-}$zixu@yeah.net}, Qiyuan Pan$^{2,3}$\footnote{Corresponding author: panqiyuan@hunnu.edu.cn}, and Jiliang Jing$^{2,3}$\footnote{jljing@hunnu.edu.cn}}
\affiliation{$^{1}$School of Science, Xi'an University of Posts and Telecommunications, Xi'an 710121, China}
\affiliation{$^{2}$ Key Laboratory of Low Dimensional Quantum Structures and Quantum Control of Ministry of Education, and Department of Physics, Hunan Normal University, Changsha, Hunan 410081, China}
\affiliation{$^{3}$Center for Gravitation and Cosmology, College of Physical Science and Technology, Yangzhou University, Yangzhou 225009, China}

\vspace*{0.2cm}
\begin{abstract}
\baselineskip=0.6 cm
\begin{center}
{\bf Abstract}
\end{center}

We employ the numerical and analytical methods to study the effects of the hyperscaling violation on the ground and excited states of holographic superconductors for $d=2,~z=2$. For both the holographic s-wave and p-wave models with the hyperscaling violation, we observe that the excited state has a lower critical temperature than the corresponding ground state, which is similar to the relativistic case, and the difference of the dimensionless critical chemical potential between the consecutive states decreases as the hyperscaling violation increases. Interestingly, as we amplify the hyperscaling violation in the s-wave model, the critical temperature of the ground state first decreases and then increases, but that of the excited states always decreases. In the p-wave model, regardless of the ground state or the excited states, the critical temperature always decreases with increasing the hyperscaling violation. In addition, we find that the hyperscaling violation affects the conductivity $\sigma$ which has $n$ peaks for the $n$-th excited state, and changes the relation in the gap frequency for the excited states in both s-wave and p-wave models.

\end{abstract}

%\keywords{AdS/CFT correspondence, Holographic superconductors, Hyperscaling violation}

\pacs{11.25.Tq, 04.70.Bw, 74.20.-z}\maketitle
\newpage
\vspace*{0.2cm}

\section{Introduction}

In modern condensed matter physics, there is still an unsolved mystery on the mechanism of high-temperature superconductivity which is not described by the usual BCS theory \cite{BCS}. Recently, it was suggested that the anti-de Sitter/conformal field theory (AdS/CFT) correspondence \cite{Maldacena,Witten,Gubser1998} can provide a powerful way to understand the high temperature superconductor systems \cite{GubserPRD78}. When the temperature of the black hole is below a critical temperature $T_{c}$, the bulk configuration becomes unstable and experiences a second order phase transition from normal state to superconducting state which brings the spontaneous U(1) symmetry breaking \cite{HartnollPRL,HartnollJHEP}. In the boundary dual CFT, these properties of the so-called holographic superconductors could be interpreted as the Cooper pair-like superconductor condensate \cite{HartnollRev,HerzogRev,HorowitzRev,CaiRev}.

The aforementioned works on the holographic superconductors only focus on the ground state. When a thermodynamic system is in equilibrium state, it corresponds to the minimum Gibbs free energy. However, the physical system is not necessarily in equilibrium, but may remain in the excited metastable state. Such metastable states manifest themselves in the jumps of magnetization, in the paramagnetic Meissner effect, superheating and supercooling phenomena, in the hysteresis, and in other peculiarities of the mesoscopic samples behavior, observed experimentally \cite{GFZharkov}. Working in the probe limit with the holographic duality, Wang \emph{et al.} studied the optical conductivity and the condensate in the excited states via the numerical method in the background of Schwarzschild-AdS black hole and found that the higher excited state has a lower critical temperature than the corresponding ground state in this relativistic case \cite{WangJHEP2020}, which was analytically confirmed in Ref. \cite{QiaoEHS} and extended to the models with backreaction in Ref. \cite{WangLLZEPJC}. Meanwhile, the conductivity $\sigma$ of each excited state has a delta function in Re[$\sigma$] and an additional pole in Im[$\sigma$] arising at the low temperature inside the gap \cite{WangJHEP2020}. As the intermediate states during the relaxation from the normal state to the ground state, the non-equilibrium condensation process of the holographic s-wave superconductor with the excited states was investigated in Ref. \cite{Liran}. More recently, the authors of Ref. \cite{OuYangliang} constructed the holographic insulator/superconductor phase transitions with excited states and studied the effects of the node number on the holographic dual model in the background of AdS soliton.

On the other hand, since many condensed matter systems do not have relativistic symmetry, it is very important to generalize the holographic superconductor models to nonrelativistic situations by using the nonrelativistic AdS/CFT correspondence. The holographic superconductors in the Lifshitz black hole with anisotropic scaling were constructed in Refs. \cite{EUTT,SinXuZhou}. Bu further observed the Lifshitz black hole geometry results in the different asymptotic behaviors of temporal and spatial components of gauge fields when compared to the previous Schwarzschild-AdS black hole \cite{BuPRD}. In the Ho\v{r}ava-Lifshitz black holes, the holographic superconductivity was a robust phenomenon associated with asymptotic AdS black holes \cite{CaiZhangHL2010} and the critical temperature of the condensation was affected by the Lifshitz exponent \cite{LuoXF}. In Ref. \cite{DHKTW}, Dong \emph{et al.} discussed the various aspects of holography for theories with hyperscaling violation from gravity side. Especially, the authors studied the effects of the hyperscaling violation on the superconducting transition temperature \cite{FanJHEP,pan} and found that the critical temperature decreases first and then increases when the hyperscaling violation $\theta$ increases for the fixed  mass square of the scalar field in the holographic s-wave model. Along this line, the holographic superconductor models of nonrelativistic situations have been studied widely \cite{WenSang,Schaposnik,Tallarita,LalaPLB2014,ZPJPLB2014,LuWuNPB,MomeniLifshitz,KEA2015,DectorNPB,JingPLB2015}.

However, it should be noted that the aforementioned studies on the holographic superconductors with hyperscaling violation were implemented on the ground state. Therefore, it is interesting to investigate the case of the excited states of holographic superconductors with hyperscaling violation. In this work, we construct a family of solutions of the holographic superconductors with the excited states in the background of the hyperscaling violation black hole, and employ the numerical method and analytical method to study the effects of the hyperscaling violation on the condensate, the critical temperature (the critical chemical potential) and the conductivity for the excited states, which shows that some properties in excited states are different from that of the ground state.

The structure of this work is organized as follows. In Sec. II, we review the black hole background with hyperscaling violation. In Sec. III, we use the numerical shooting method and analytical Sturm-Liouville method to investigate the excited states of holographic s-wave superconductors with hyperscaling violation. In Sec. IV, we study the p-wave case. We summarize our results in the last section.

\section{Black hole background with hyperscaling violation}

In the probe limit, we will construct the holographic superconductors with excited states in the background of the hyperscaling violation black hole \cite{DHKTW}
\begin{eqnarray}\label{Soliton}
ds_{d+2}^2=r^{-2(d-\theta)/d}\left[-r^{-2(z-1)}f(r)dt^2+\frac{dr^2}{f(r)}+dx_{i}^{2}\right],
\end{eqnarray}
where $f(r)=1-\left(\frac{r}{r_{+}}\right)^{d+z-\theta}$ with the radius of the event horizon $r_{+}$, dynamical exponent $z$ and hyperscaling violation exponent $\theta$. Obviously, this metric will reduce to the pure AdS case when $\theta=0$ and $z=1$, and describe the pure Lifshitz case when $\theta=0$ and $z\neq1$. The Hawking temperature of the black hole is given by
\begin{eqnarray}
T=\frac{d+z-\theta}{4\pi r_{+}^{z}},
\end{eqnarray}
which is interpreted as the temperature of the dual system from the AdS/CFT correspondence.

At the asymptotic boundary $r\rightarrow0$, from (\ref{Soliton}) we obtain the most general metric
\begin{eqnarray}
ds_{d+2}^2=r^{-2(d-\theta)/d}\left[-r^{-2(z-1)}dt^2+dr^2+dx_{i}^{2}\right],
\end{eqnarray}
which is spatially homogeneous and covariant under the scale transformations
\begin{eqnarray}
r\rightarrow\alpha r,~~t\rightarrow\alpha^{z}t,~~~x_{i}\rightarrow\alpha
x_{i},~~ds\rightarrow\alpha^{\theta/d}ds,
\end{eqnarray}
where $\alpha$ is a real positive number. It is interesting to note that the proper distance of the spacetime transforms non-trivially under scale transformations with the exponent $\theta$, which implies a hyperscaling violation in the dual field theory \cite{DHKTW,ACYJHEP}.

Introducing the coordinate transformation $u=r/r_{+}$, we can rewrite the metric (\ref{Soliton}) into
\begin{eqnarray}
ds_{d+2}^2=(r_{+}u)^{-2(d-\theta)/d}\left[-(r_{+}u)^{-2(z-1)}f(u)dt^2+\frac{r_{+}^{2}}{f(u)}du^2+dx_{i}^{2}\right],
\end{eqnarray}
with the metric coefficient $f(u)=1-u^{d+z-\theta}$. We point out that, according to the null energy conditions, the hyperscaling violation exponent $\theta$ and dynamical exponent $z$ satisfy the following conditions \cite{DHKTW,FanJHEP,pan}
\begin{eqnarray}
d>\theta\geq0,~~z\geq1+\frac{\theta}{d}.
\end{eqnarray}
In order to consider the effects of hyperscaling violation $\theta$ on the holographic superconductors with excited states for clarity, we will focus on $d=2,~z=2$ and $\theta\geq0$ in this work, just as in Ref. \cite{pan}.

\section{Excited states of the s-wave superconductors with hyperscaling violation}

\subsection{Condensates of the scalar field}

In the probe limit, we will consider a gauge field coupled with a scalar field via the action
\begin{eqnarray}\label{System}
S=\int d^{4}x\sqrt{-g}\left(
-\frac{1}{4}F_{\mu\nu}F^{\mu\nu}-|\nabla_{\mu}\psi - iA_{\mu}\psi|^{2}
-m^2|\psi|^2 \right) \ ,
\end{eqnarray}
where $m$ is the mass of the scalar field $\psi$. By adopting the ansatz for the matter fields as $\psi=\psi(u)$ and $A=\phi(u) dt$, we can get the equations of motion
\begin{eqnarray}
&&\psi^{\prime\prime}+\left(\frac{f^\prime}{f}-
\frac{3-\theta}{u}\right)\psi^\prime
+r_{+}^{4}\left(\frac{u^{2}\phi^2}{f^2}-\frac{m^2r_{+}^{\theta-4}}{u^{2-\theta}f}\right)\psi=0,
\label{BHPsiu}
\end{eqnarray}
\begin{eqnarray}
\phi^{\prime\prime}+\frac{1}{u}\phi^\prime-\frac{2r_{+}^{\theta}\psi^{2}}{u^{2-\theta}f}\phi=0,\label{BHPhiu}
\end{eqnarray}
where the prime denotes the derivative with respect to $u$ and this notation will be used in the following discussion. It should be noted that, when $\theta\rightarrow0$, Eqs. (\ref{BHPsiu}) and  (\ref{BHPhiu}) reduce to the ones in the standard holographic s-wave superconductors with $z=2$ Lifshitz scaling discussed in \cite{BuPRD}.

To obtain the solutions in the superconducting phase, i.e., $\psi(u)\neq0$, we have to specify the boundary conditions for the matter fields. At the event horizon $u=1$, we impose the boundary conditions by requiring that the scalar field  $\psi$ is regular and the gauge field satisfies $\phi(1)=0$. Near the asymptotic boundary $u\rightarrow0$, the solutions behave as
\begin{eqnarray}\label{PsiInfinity}
\psi=\left\{
\begin{array}{rl}
&\psi_{4-\Delta}r_{+}^{4-\Delta}u^{4-\Delta}+\psi_{\Delta}r_{+}^{\Delta}u^{\Delta}\,,~~~{\rm
where}\ \Delta=2+\sqrt{4+m^{2}r_{+}^{\theta}}\ {\rm for}\ \theta=0,
\\ \\ &\psi_{0}+\psi_{\Delta}r_{+}^{\Delta}u^{\Delta}  \,,~~
\quad\quad\quad\quad\quad\quad {\rm where}\ \Delta=4-\theta\ {\rm
for}\ 0<\theta<2,
\end{array}\right.
\end{eqnarray}
\begin{eqnarray}
\phi=\rho+\mu\ln u,
\end{eqnarray}
where $\Delta$ is the characteristic exponent, $\mu$ and $\rho$ are interpreted as the chemical potential and charge density in the dual field theory, respectively. In order to investigate the condensate $\psi_{\Delta}=\langle O\rangle$, we will impose boundary conditions $\psi_{4-\Delta}=0$ and $\psi_{0}=0$, just as in Refs. \cite{BuPRD,FanJHEP,FanJHEP1,FanPRD}. Moreover, we will set $m^{2}r_{+}^{\theta}=-3$ for concreteness since the other choices will not qualitatively modify our results.

It is interesting to note that, in the following discussion, we can build the invariant and dimensionless quantities by using the scaling transformations
\begin{eqnarray}
&&\psi\rightarrow\beta^{-\frac{\theta}{2}}\psi,\hspace{0.5cm}\phi\rightarrow\beta^{-2}\phi,\nonumber\\
&&\psi_{\Delta}\rightarrow\beta^{-(\Delta+\frac{\theta}{2})}\psi_{\Delta},\hspace{0.5cm}\mu\rightarrow\beta^{-2}\mu,\hspace{0.5cm}
\end{eqnarray}
where $\beta$ is a positive number.

\subsubsection{Numerical analysis}

\begin{figure}[ht]
\includegraphics[scale=0.95]{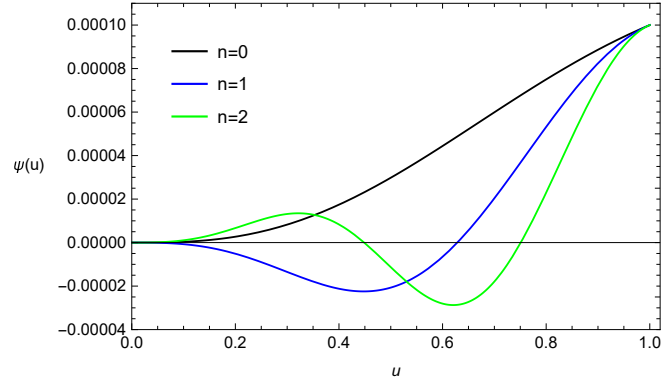}\\ \vspace{0.2cm}
\caption{\label{SNodes} (Color online.) The distribution of the scalar field $\psi(u)$ as a function of $u$ for the condensate $\langle O\rangle$ with the hyperscaling violation $\theta = 1.0$. The black, blue and green lines denote the ground state, first and second excited states, respectively.}
\end{figure}

We first use the shooting method \cite{HartnollPRL,HartnollJHEP} to solve numerically the equations of motion (\ref{BHPsiu}) and (\ref{BHPhiu}), which gives us the solutions of the ground and excited states. In Fig. \ref{SNodes}, we show the distribution of the scalar field $\psi(u)$ as a function of $u$ for the condensate $\langle O\rangle$ with the hyperscaling violation $\theta=1.0$. The black line has no intersecting point with $\psi(u)=0$, which denotes the ground state with the number of the nodes $n=0$. And the blue line has one intersecting point with $\psi(u)=0$ while the green line has two intersecting points, corresponding to the first $(n = 1)$ and second $(n = 2)$ excited states, respectively. Similarly, the $n$-th excited state has exactly $n$ nodes.

\begin{figure}[H]
\includegraphics[scale=0.6]{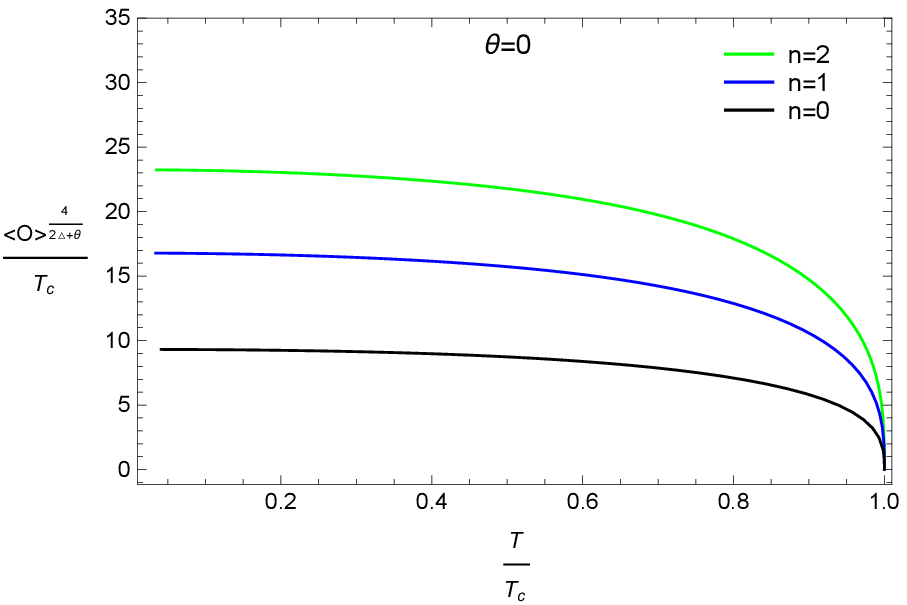}\hspace{0.2cm}%
\includegraphics[scale=0.6]{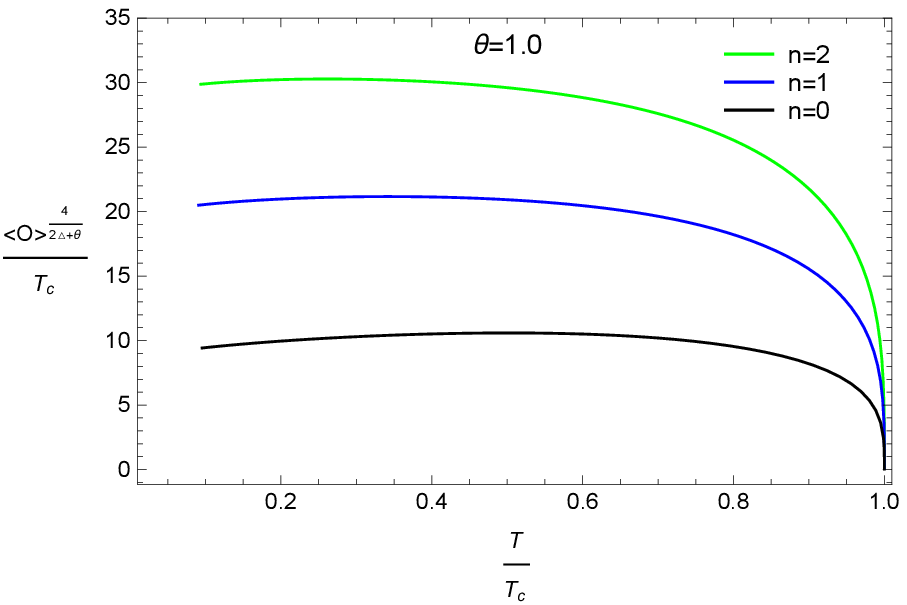}\vspace{0.2cm}%
\includegraphics[scale=0.6]{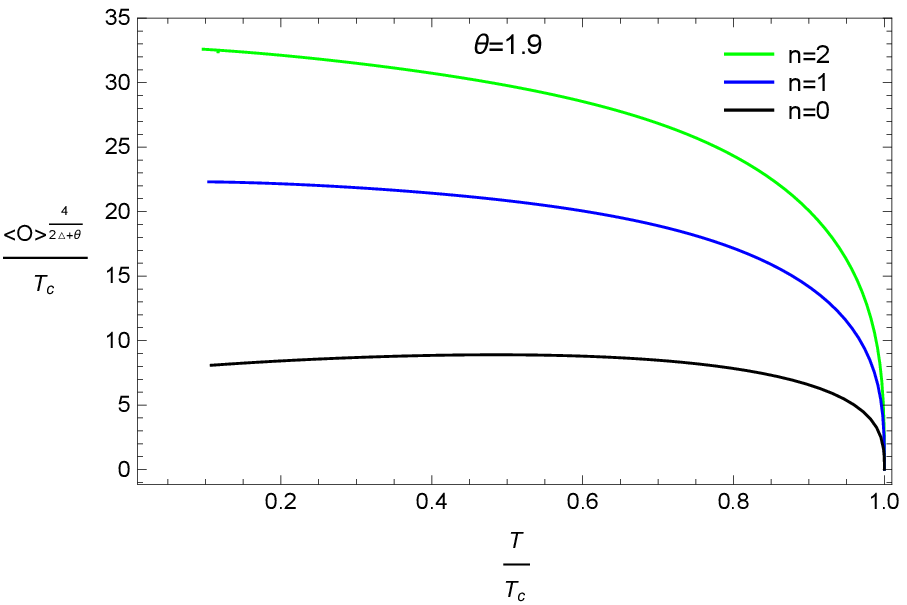}\\ \vspace{0.1cm}
\caption{\label{SwaveCondensate0}(Color online.) The condensate of the scalar operator $O$ with $\theta=0$ (left panel), $\theta=1.0$ (middle panel) and $\theta=1.9$ (right panel) as a function of temperature. In each panel, the black, blue and green lines denote the ground, first and second excited states, respectively.}
\end{figure}

We will discuss the effects of the hyperscaling violation on the condensation of the scalar operators. In Fig. \ref{SwaveCondensate0}, we plot the condensate of the scalar operator $O$ as a function of temperature with different values of the hyperscaling violation $\theta$. In each panel, as the temperature $T\rightarrow 0$, the condensate quickly saturates a fixed value which increases with the increase of $n$. Fitting these condensation curves for the operator $O$ with $\theta=0,~1.0$ and $1.9$ for small condensate, we obtain
\begin{eqnarray}
\langle O_{0}\rangle\approx
\left\{
\begin{array}{rl}
44.566(T_{c}^{(0)})^{3/2}(1-T/T_{c}^{(0)})^{1/2}, \\ \\
112.485(T_{c}^{(1)})^{3/2}(1-T/T_{c}^{(1)})^{1/2} \ , &  \quad  {\rm for}\ \theta=0 ,\\ \\ 186.325(T_{c}^{(2)})^{3/2}(1-T/T_{c}^{(2)})^{1/2},
\end{array}\right.
\end{eqnarray}
\begin{eqnarray}
\langle O_{1.0}\rangle\approx
\left\{
\begin{array}{rl}
129.707(T_{c}^{(0)})^{7/4}(1-T/T_{c}^{(0)})^{1/2}, \\ \\
404.323(T_{c}^{(1)})^{7/4}(1-T/T_{c}^{(1)})^{1/2} \ , &  \quad  {\rm for}\ \theta=1.0 ,\\ \\ 731.296(T_{c}^{(2)})^{7/4}(1-T/T_{c}^{(2)})^{1/2},
\end{array}\right.
\end{eqnarray}
\begin{eqnarray}
\langle O_{1.9}\rangle\approx
\left\{
\begin{array}{rl}
57.557(T_{c}^{(0)})^{61/40}(1-T/T_{c}^{(0)})^{1/2}, \\ \\
187.372(T_{c}^{(1)})^{61/40}(1-T/T_{c}^{(1)})^{1/2} \ , &  \quad  {\rm for}\ \theta=1.9,\\ \\ 320.946(T_{c}^{(2)})^{61/40}(1-T/T_{c}^{(2)})^{1/2},
\end{array}\right.
\end{eqnarray}
where $T_{c}^{(0)}$, $T_{c}^{(1)}$ and $T_{c}^{(2)}$ are presented in Table \ref{NMTcTable}, corresponding to the critical temperatures for $n=0$, $n=1$, and $n=2$, respectively. For all cases considered here, we see that for small condensate there exists a square root behavior
\begin{eqnarray}
\langle O\rangle\sim(1-T/T_{c}^{(n)})^{1/2},
\end{eqnarray}
which shows that the phase transition of the holographic s-wave superconductors with the hyperscaling violation in the excited states belongs to the second order and the critical exponent of the system takes the mean-field value 1/2 for all the values of $\theta$.

\begin{table}[H]
\begin{center}
\caption{\label{NMTcTable} The critical temperature $T_{c}$ obtained via the shooting method for the scalar operator $O$ with different values of the hyperscaling violation $\theta$.}
\begin{tabular}{c| c c c c c c c}
         \hline \hline
$\theta$ & 0 & 0.5 & 0.6 & 0.7 & 1.0 & 1.5 & 1.9
        \\
        \hline
$n=0$~~&~~$0.0351935\mu$~~&~~$0.0330971\mu$~~&~~$0.0330311\mu$~~&~~$0.0330385\mu$~~&~~$0.0334913\mu$~~&~~$0.0361219\mu$~~&~~$0.0420384\mu$~
          \\
          \hline
$n=1$~~&~~$0.0168623\mu$~~&~~$0.0150635\mu$~~&~~$0.0148259\mu$~~&~~$0.0146057\mu$~~&~~$0.0140200\mu$~~&~~$0.0132050\mu$~~&~~$0.0126560\mu$~
          \\
          \hline
$n=2$~~&~~$0.0110130\mu$~~&~~$0.0098142\mu$~~&~~$0.0096401\mu$~~&~~$0.0094750\mu$~~&~~$0.0090176\mu$~~&~~$0.0083311\mu$~~&~~$0.0078231\mu$~
          \\
          \hline
$n=3$~~&~~$0.0081628\mu$~~&~~$0.0072880\mu$~~&~~$0.0071543\mu$~~&~~$0.0070263\mu$~~&~~$0.0066657\mu$~~&~~$0.0061115\mu$~~&~~$0.0056934\mu$~
          \\
          \hline
$n=4$~~&~~$0.0064805\mu$~~&~~$0.0057989\mu$~~&~~$0.0056911\mu$~~&~~$0.0055873\mu$~~&~~$0.0052919\mu$~~&~~$0.0048321\mu$~~&~~$0.0044825\mu$~
          \\
          \hline
$n=5$~~&~~$0.0053716\mu$~~&~~$0.0048161\mu$~~&~~$0.0047261\mu$~~&~~$0.0046389\mu$~~&~~$0.0043894\mu$~~&~~$0.0039979\mu$~~&~~$0.0036989\mu$~
          \\
        \hline \hline
\end{tabular}
\end{center}
\end{table}

\begin{figure}[htbp]
\includegraphics[scale=0.95]{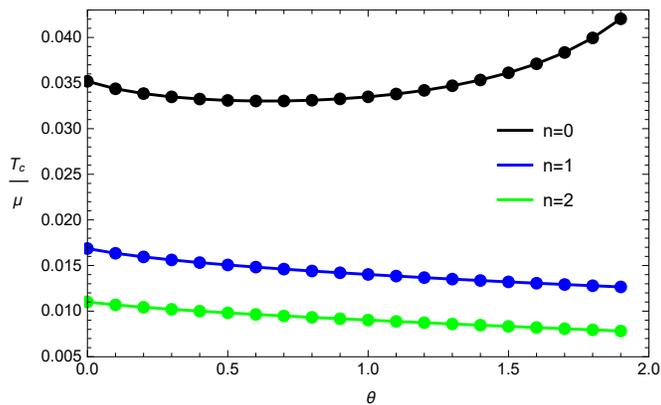}\\ \vspace{0.2cm}
\caption{\label{TcCondensate} (Color online.) The critical temperature $T_{c}$ as a function of the hyperscaling violation $\theta$ in the holographic s-wave superconductors. The three lines from top to bottom correspond to the ground ($n=0$, black), first ($n=1$, blue) and second($n=2$, green) excited states, respectively.}
\end{figure}

In order to study the effect of the hyperscaling violation on the critical temperature, we give the critical temperature with the first six lowest-lying modes for different values of $\theta$ in Table \ref{NMTcTable}. Obviously, for the fixed $\theta$, the critical temperature $T_{c}$ decreases as $n$ increases. Interestingly, as we amplify $\theta$, the critical temperature of the ground state first decreases and then increases. However, the critical temperature always decreases with the increase of $\theta$ in the excited states, which is obviously different from the ground state. We can also find clearly these properties in Fig. \ref{TcCondensate}.

\begin{table}[h]
\caption{\label{Schepotential} The critical chemical potential $\mu_{c}$ obtained by the shooting method for the scalar operator $O$ from the ground state to the sixth excited state with various values of the hyperscaling violation $\theta$.}
\begin{tabular}{c c c c c c c c}
         \hline
$n$ & 0 & 1 & 2 & 3 & 4 & 5 & 6
        \\
        \hline
~~~~$\theta=0$~~~~&~~~~~$9.045$~~~~~&~~~~~$18.877$~~~~~&~~~~~$28.903$~~~~~&~~~~~$38.995$~~~~
&~~~~~$49.118$~~~~&~~~~~$59.258$~~~~&~~~~~$69.408$~~~~
          \\
~~~~$\theta=1.0$~~~~&~~~~~$7.128$~~~~~&~~~~~$17.028$~~~~~&~~~~~$26.474$~~~~~&~~~~~$35.815$~~~~
&~~~~~$45.113$~~~~&~~~~~$54.389$~~~~&~~~~~$63.651$~~~~
          \\
~~~~$\theta=1.9$~~~~&~~~~~$3.975$~~~~~&~~~~~$13.204$~~~~~&~~~~~$21.361$~~~~~&~~~~~$29.352$~~~~
&~~~~~$37.281$~~~~&~~~~~$45.179$~~~~&~~~~~$53.060$~~~~
          \\
        \hline
\end{tabular}
\end{table}

\begin{figure}[h]
\includegraphics[scale=0.6]{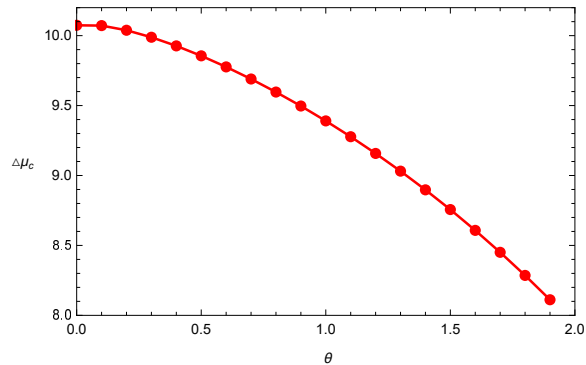}\\ \vspace{0.2cm}
\caption{\label{SChemicalPotential0} (Color online.) The difference of the  dimensionless critical chemical potential between the consecutive states $\Delta\mu_{c}$ as a function of the hyperscaling violation $\theta$ by the shooting method.}
\end{figure}

On the other hand, recent efforts showed that the difference of the dimensionless critical chemical potential between the consecutive states is around $5$ \cite{WangJHEP2020,QiaoEHS}. It is of great interest to consider the effect of the hyperscaling violation on this difference. In Table \ref{Schepotential}, we present the results of $\mu_{c}$ from the ground state to the sixth excited state by the shooting method, which shows that the critical chemical potential $\mu_{c}$ increases as the number of nodes $n$ increases for the fixed $\theta$. Thus, the excited state has a higher critical potential than the ground state, which indicates that the ground state firstly condensates when increasing the chemical potential. The relation between $\mu_{c}$ and $n$ can be fitted as
\begin{eqnarray}
\mu_{c}\approx
\left\{
\begin{array}{rl}
10.074n+8.865, \quad  {\rm for}\ \theta=0, \\ \\
9.390n+9.486, \quad  {\rm for}\ \theta=1.0, \\ \\
8.112n+4.724, \quad  {\rm for}\ \theta=1.9.
\end{array}\right.
\end{eqnarray}
Obviously, the hyperscaling violation affects the difference of the dimensionless critical chemical potential. In Fig. \ref{SChemicalPotential0}, we exhibit the difference of the dimensionless critical chemical potential between the consecutive states $\Delta\mu_{c}$ as a function of the hyperscaling violation $\theta$ by the shooting method, which tells us that the difference of the dimensionless critical chemical potential between the consecutive states decreases as $\theta$ increases.

\subsubsection{Analytical investigation}

Now we are in a position to study analytically the properties of the excited states of the s-wave superconductor models with the hyperscaling violation by using the Sturm-Liouville method \cite{Siopsis,SiopsisJ}, which can be used to back up the numerical calculations.

At the critical temperature $T_{c}$, the scalar field $\psi=0$. So, near the critical point, Eq. (\ref{BHPhiu})
reduces to $\phi^{\prime\prime}+\frac{1}{u}\phi^{\prime}=0$, which has a solution
\begin{eqnarray}
\phi(u)=\lambda r_{+c}^{-2}\ln u,\label{PhiSolution0}
\end{eqnarray}
where we have introduced $\lambda=\mu r_{+c}^{2}$ with the radius of the horizon at the critical point $r_{+c}$. In order to match the boundary behavior (\ref{PsiInfinity}) for $\psi$, we define a trial function $F(u)$ which satisfies
\begin{eqnarray}
\psi(u)\sim\langle O\rangle r_{+}^{\Delta}u^{\Delta}F(u),
\end{eqnarray}
with the boundary conditions $F(0)=1$ and $F'(0)=0$. Thus, we can get the equation of motion
\begin{eqnarray}
(TF^{\prime})^{\prime}+T\left(U+\lambda^2V\right)F=0,
\end{eqnarray}
where
\begin{eqnarray}\label{BFEoM1}
T(u)=u^{2\Delta+\theta-3}f,~~U(u)=\frac{\Delta
}{u}\left(\frac{f'}{f}+\frac{\Delta+\theta-4}{u}\right)-\frac{m^2r_{+}^{\theta}}{u^{2-\theta}f},~~V(u)=\left(\frac{u\ln
u}{f}\right)^{2}.
\end{eqnarray}
Following the Sturm-Liouville eigenvalue problem \cite{Gelfand-Fomin}, we can obtain the eigenvalues of $\lambda^{2}$ from variation of the expression
\begin{eqnarray}\label{lambdaeigenvalue1}
\lambda^{2}=\left(\mu r_{+c}^{2}\right)^{2}=\frac{\int^{1}_{0}T\left(F'^{2}-UF^{2}\right)du}{\int^{1}_{0}TVF^2du}.
\end{eqnarray}

In order to analytically investigate the excited states of the holographic superconductors with the hyperscaling violation, we choose the ninth order of $u$ in the trial function $F(u)$, i.e., $F(u)=1-\sum_{k=2}^{k=9}a_{k}u^{k}$ for the operator $O$, where $a_{k}$ is a constant \cite{QiaoEHS,OuYangliang}. As an example, we set the hyperscaling violation $\theta=1.5$ and obtain the critical chemical potential $\mu_{c}$ from ground state to the fifth excited state which can be seen in Table \ref{SWaveO1Table1} by using Eq. (\ref{lambdaeigenvalue1}). With the expression
\begin{eqnarray}
T_{c}=\frac{4-\theta}{4\pi\lambda_{ext}}\mu,
\end{eqnarray}
where $\lambda_{ext}$ is the extremal values of Eq. (\ref{lambdaeigenvalue1}), we also give the critical temperature $T_{c}$ from the ground state to the fifth excited state in Table \ref{TcSWave}. Comparing with numerical results in Tables \ref{SWaveO1Table1} and \ref{TcSWave}, we observe that our analytical results are in very good agreement with the numerical calculation. This tiny error implies that we can use the Sturm-Liouville method to back up the numerical finding as shown in the holographic s-wave superconductors with the hyperscaling violation that, for the fixed $\theta$, the critical chemical potential $\mu_{c}$ (the critical temperature $T_{c}$) increases (decreases) as the number of nodes $n$ increases.

\begin{table}[H]
\begin{center}
\caption{\label{SWaveO1Table1}
The dimensionless critical chemical potential $\mu_{c} r_{+c}^{2}$ for the operator $O$ and the corresponding value of $a_{k}$ for the trial function $F(u)=1-\sum_{k=2}^{k=9}a_{k}u^{k}$. The results of $\mu_{c} r_{+c}^{2}$ are obtained analytically by the Sturm-Liouville method (left column) and numerically by the shooting method (right column) with $\theta=1.5$ from the ground state to the fifth excited state.}
\begin{tabular}{c|c| c c c c c c c c c}
\hline
 $n$&$\mu_{c} r_{+c}^{2}$~&~$a_{2}$~&~$a_{3}$~&~$a_{4}$~&~$a_{5}$~&~$a_{6}$~&~$a_{7}$~&~$a_{8}$~&~$a_{9}$  \\
\hline
$0$&~5.508~~5.508~&~1.729~&~-6.429~&~21.268~&~-43.877~&~55.715~&~-43.079~&~18.670~&~-3.481\\
\hline
$1$&~15.066~~15.066~&~1.708~&~-1.607~&~40.972~&~-125.901~&~162.304~&~-108.068~&~37.014~&~-5.193 \\
\hline
$2$&~23.880~~23.880~&~6.651~&~-56.371~&~426.422~&~-1324.827~&~2022.943~&~-1645.905~&~689.211~&~-117.282 \\
\hline
$3$&~32.555~~32.552~&~15.170~&~-201.677~&~1635.370~&~-5910.339~&~10898.044~&~-10847.608~&~5579.912~&~-1167.750 \\
\hline
$4$&~41.413~~41.171~&~-225.740~&~2437.863~&~-9723.630~&~19133.645~&~-19622.891~&~9473.421~&~-1032.869~&~-439.001 \\
\hline
$5$&~50.557~~49.763~&~-1793.978~&~22390.945~&~-112761.897~&~301337.743~&~-465189.986~&~417160.143~&~-202070.989~&~40929.526 \\
\hline
\end{tabular}
\end{center}
\end{table}

\begin{table}[ht]
\caption{\label{TcSWave}
The critical temperature $T_{c}$ obtained numerically by the shooting method ($T_{cn}$) and analytically by the Strum-Liouville method $T_{ca}$ for the scalar operator $O$ with $\theta=1.5$ from the ground state to the fifth excited state.}
\begin{tabular}{c c c c c c c c}
         \hline
$n$ & 0 & 1 & 2 & 3 & 4 & 5
        \\
        \hline
~~$T_{cn}$~~&~~~~~$0.0361219\mu$~~~~~&~~~~~$0.0132050\mu$~~~~~&~~~~~$0.0083311\mu$~~~~~&~~~~~$0.0061115\mu$~~~~
&~~~~~$0.0048321\mu$~~~~&~~~~~$0.0039979\mu$
          \\
~~$T_{ca}$~~&~~~~~$0.0361219\mu$~~~~~&~~~~~$0.0132051\mu$~~~~~&~~~~~$0.0083311\mu$~~~~~&~~~~~$0.0061110\mu$~~~~
&~~~~~$0.0048039\mu$~~~~&~~~~~$0.0039350\mu$
          \\
        \hline
\end{tabular}
\end{table}

Furthermore, we will study the critical phenomena of the holographic s-wave superconductors with the hyperscaling violation. As $T\rightarrow T_{c}$, the condensate of the scalar operator $O$ is so small. Therefore, we can expand  $\phi(u)$
in small $\langle O\rangle$ as
\begin{eqnarray}\label{PhiExpandNearTc}
\phi(u)=\lambda r_{+}^{-2}\ln
u+r_{+}^{2(\Delta-1)+\theta}\langle O\rangle^{2}\chi(u)+\cdot\cdot\cdot,
\end{eqnarray}
with the boundary conditions $\chi(1)=0$ and $\chi'(1)=0$ at the event horizon \cite{Siopsis}, which leads to the equation of motion for $\chi(u)$
\begin{eqnarray}
(u\chi^\prime)^\prime-\frac{2\lambda u^{2\Delta+\theta-1}F^{2}\ln
u}{f}=0.
\end{eqnarray}
Making integration for both sides of the above equation, we get
\begin{eqnarray}
(u\chi^\prime)|_{u\rightarrow
0}=\lambda\mathcal{A}=-\lambda\int^{1}_{0}\frac{2u^{2\Delta+\theta-1}F^{2}\ln
u}{f}du.
\end{eqnarray}

According to Eqs. (\ref{PhiSolution0}) and (\ref{PhiExpandNearTc}), near
$u\rightarrow0$ we have
\begin{eqnarray}
\mu=\lambda
r_{+}^{-2}+r_{+}^{2(\Delta-1)+\theta}\langle O\rangle^{2}(u\chi^\prime)|_{u\rightarrow 0},
\end{eqnarray}
which gives us
\begin{eqnarray}\label{OExp}
\langle O\rangle=\frac{1}{\sqrt{\mathcal{A}}}\left(\frac{4\pi
T_{c}}{4-\theta}\right)^{\frac{2\Delta+\theta}{4}}\left(1-\frac{T}{T_c}\right)^{\frac{1}{2}}.
\end{eqnarray}
It is clearly shown that the expression (\ref{OExp}) is valid for all cases considered here. For example, for the case of $\theta=1.5$, we obtain
\begin{eqnarray}
\langle O\rangle\approx
\left\{
\begin{array}{rl}
60.2(T_{c}^{(0)})^{13/8}(1-T/T_{c}^{(0)})^{1/2}   \ , &  \quad {\rm ground~state}, \\ \\
161.3(T_{c}^{(1)})^{13/8}(1-T/T_{c}^{(1)})^{1/2}   \ , &  \quad {\rm 1st~excited~state}, \\ \\ 270.5(T_{c}^{(2)})^{13/8}(1-T/T_{c}^{(2)})^{1/2}   \ , &  \quad {\rm 2nd~excited~state},
\end{array}\right.
\end{eqnarray}
which agrees well with the numerical calculation by the shooting method. Here the critical temperature $T_{c}^{(0)}$, $T_{c}^{(1)}$ and $T_{c}^{(2)}$, which correspond to the ground, first and second excited states, are
given in Table \ref{TcSWave}. Thus, the holographic s-wave superconductor phase transition with the hyperscaling violation in the excited states is always the second order with the mean-field critical exponent $1/2$, and the hyperscaling violation $\theta$ and the number of nodes $n$ do not affect the order of phase transition.

\subsection{Conductivity}

In order to discuss the conductivity in the holographic s-wave superconductor models with the hyperscaling violation
from the ground state to the excited states, we assume that the perturbed Maxwell field has a form  $\delta A_{x}=A_{x}(u)e^{-i\omega t}dx$. So the equation of motion for $A_{x}$ is given by
\begin{eqnarray}
A_{x}^{\prime\prime}+\left(\frac{f^\prime}{f}-\frac{1}{u}\right)A_{x}^\prime
+\left(\frac{r_{+}^{4}\omega^2u^{2}}{f^2}-\frac{2r_{+}^{\theta}\psi^{2}}{u^{2-\theta}f}\right)A_{x}=0.
\label{ConductivityEquationSWave}
\end{eqnarray}
Near the horizon, the ingoing wave boundary condition for different hyperscaling violation exponents is $A_{x}(u)\sim (1-u)^{-i\omega/(4\pi T)}$. And in the asymptotic AdS region ($u\rightarrow 0$), the behavior can be expressed as
\begin{eqnarray}
A_{x}=A_{x}^{(0)}+A_{x}^{(1)}r_{+}^{2}u^{2}.
\end{eqnarray}
From the AdS/CFT correspondence, the conductivity can be written as
\begin{eqnarray}
\sigma=\frac{A_{x}^{(1)}}{i\omega A_{x}^{(0)}}\ .
\end{eqnarray}
We numerically solve the Maxwell equation (\ref{ConductivityEquationSWave}) to obtain the conductivity for different values of the hyperscaling violation $\theta$.

\begin{figure}[H]
\includegraphics[scale=0.48]{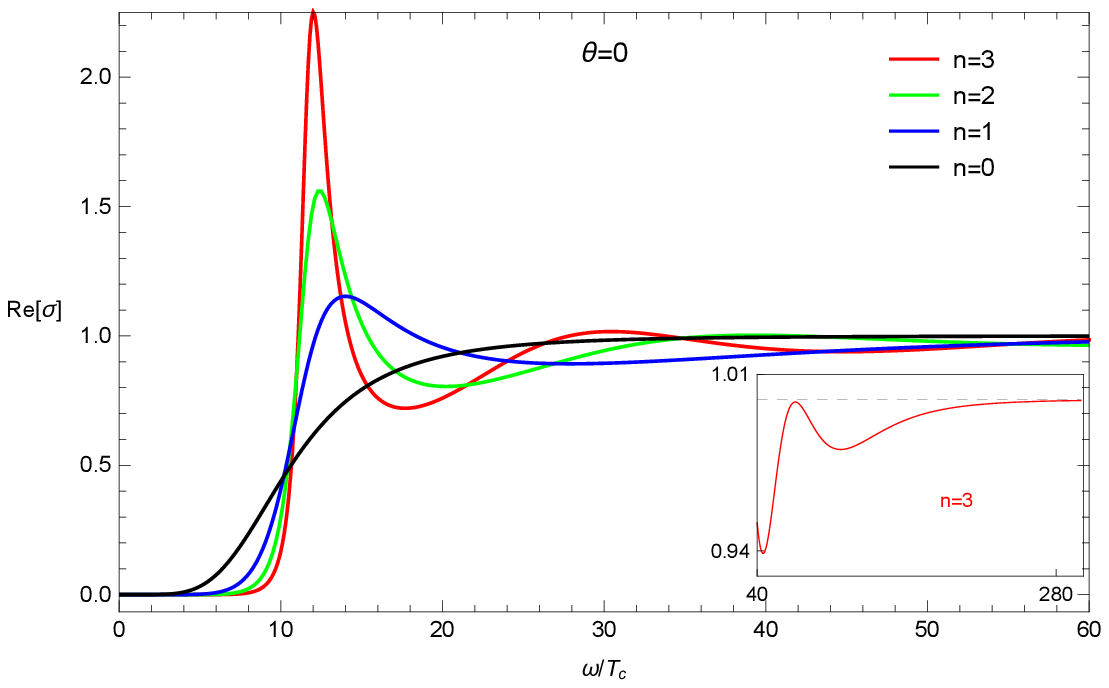}\hspace{0.2cm}%
\includegraphics[scale=0.46]{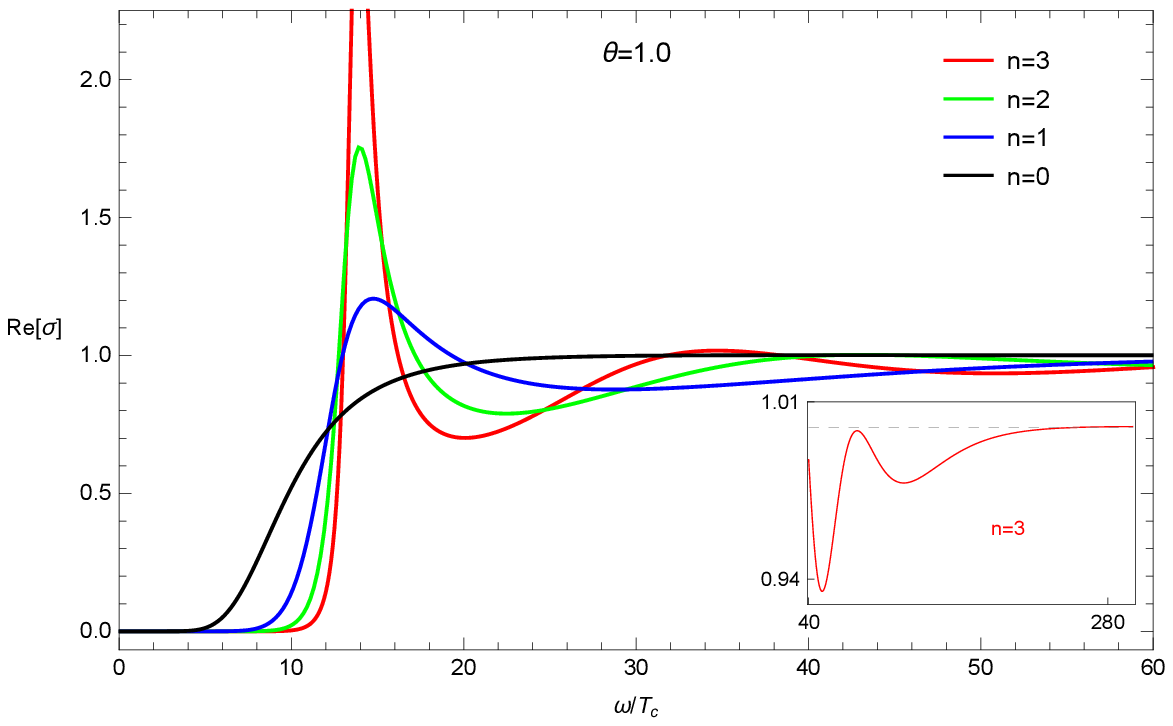}\hspace{0.2cm}%
\includegraphics[scale=0.46]{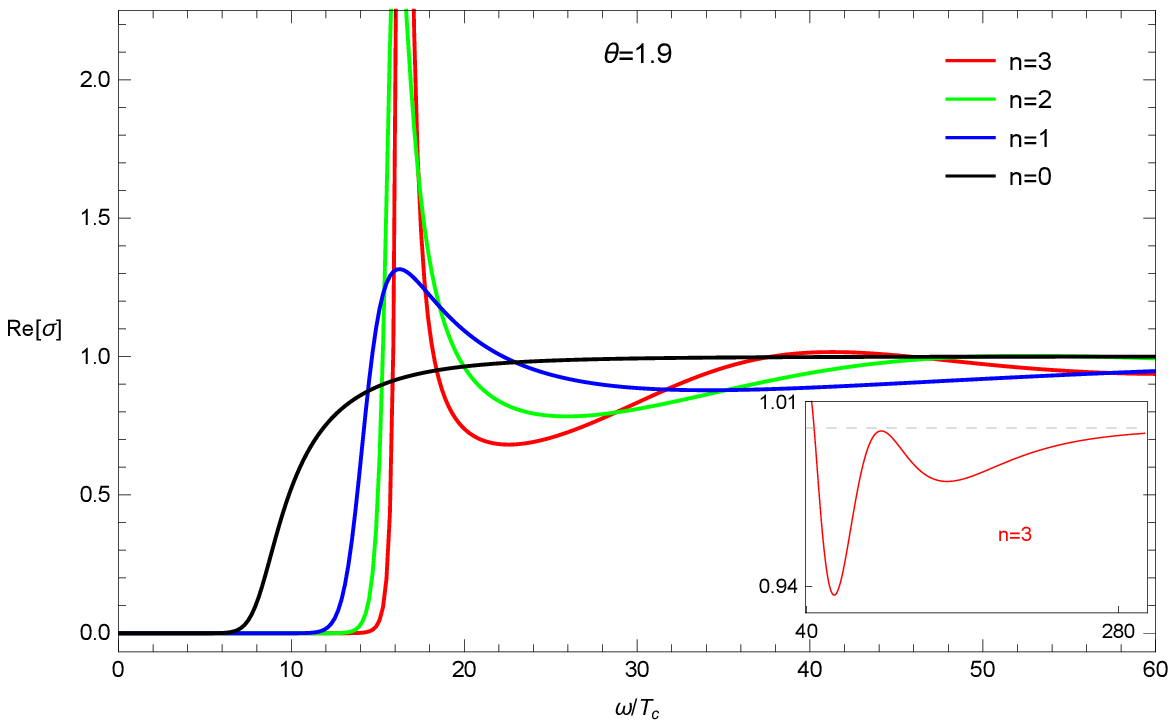}\\ \vspace{0.2cm}
\includegraphics[scale=0.45]{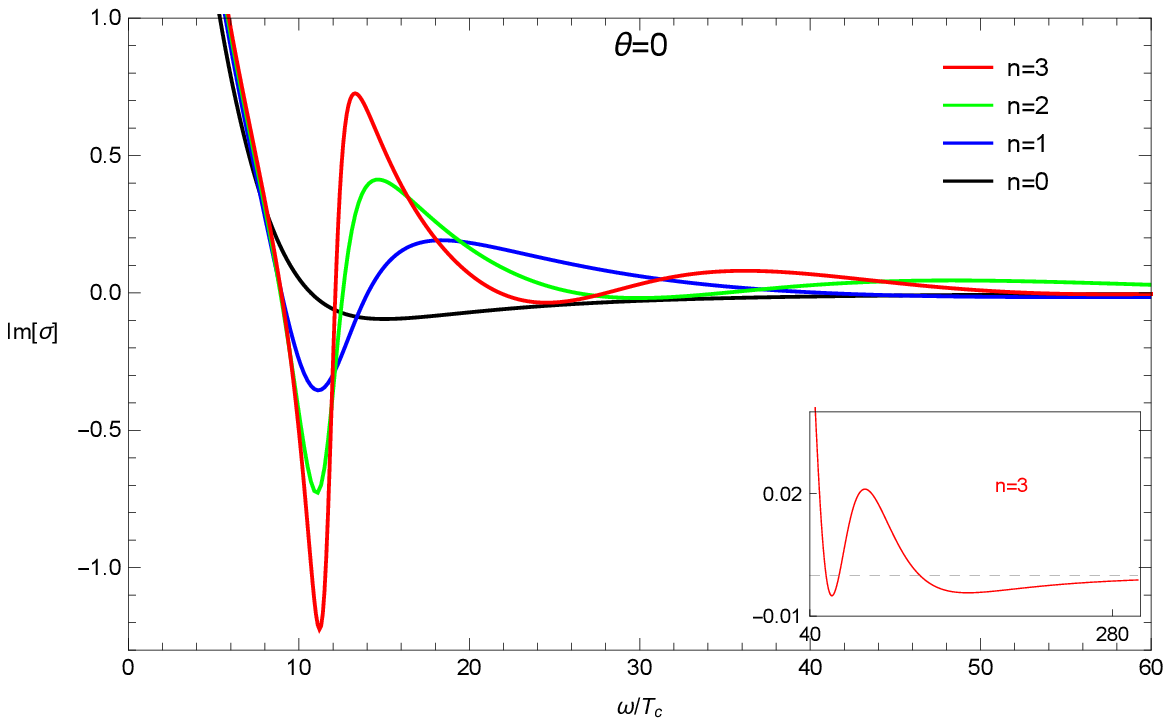}\hspace{0.2cm}%
\includegraphics[scale=0.45]{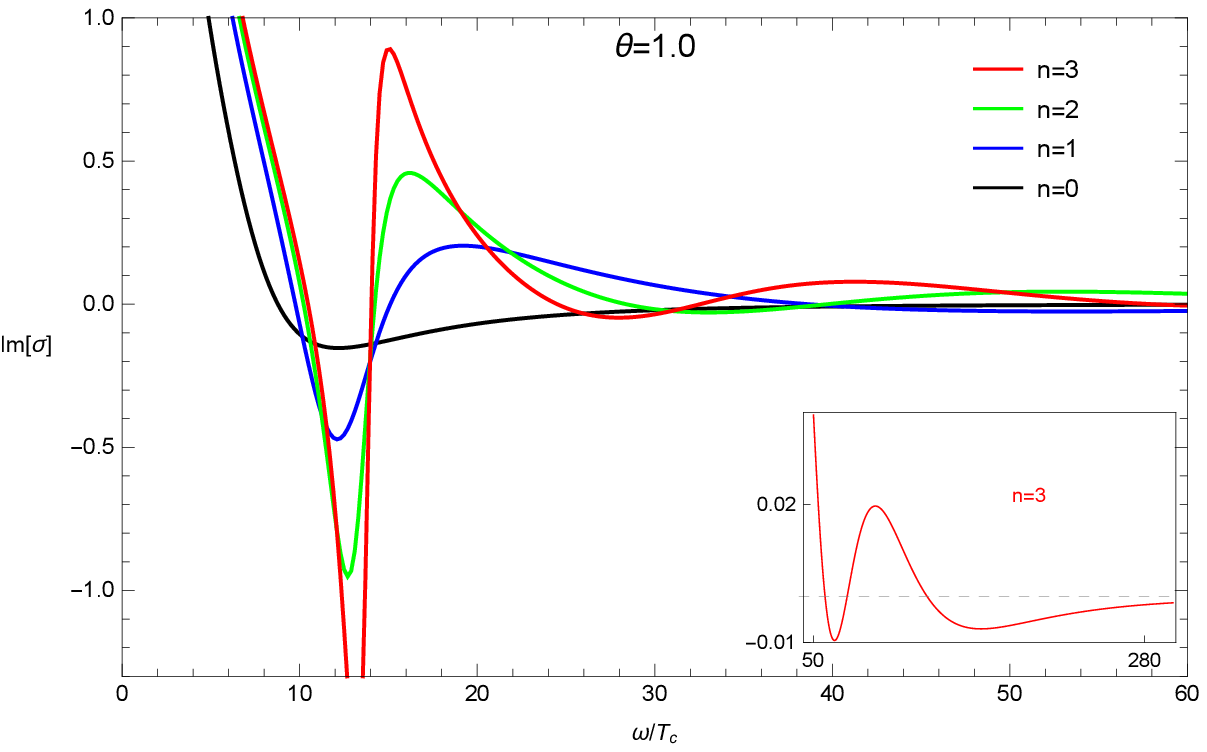}\hspace{0.2cm}%
\includegraphics[scale=0.46]{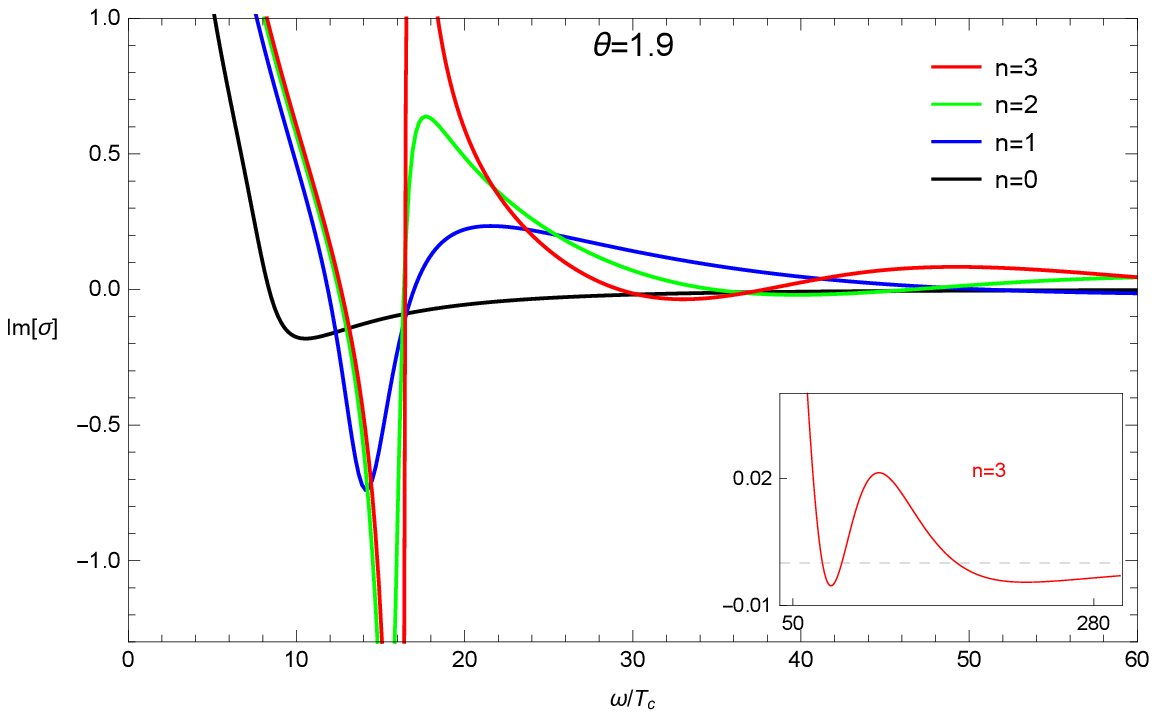}\\ \vspace{0.2cm}
\caption{\label{Conductivity0}(Color online.) The conductivity of the holographic s-wave superconductors for different values of the hyperscaling violation. In each panel, the black, blue, green and red lines denote the ground state, first, second and third excited states, respectively.}
\end{figure}

In Fig. \ref{Conductivity0}, in the case of $m^{2}r_{+}^{\theta}=-3$, we plot the real part and imaginary parts of the conductivity of the holographic s-wave models with the hyperscaling violation for $\theta=0$, $1.0$ and $1.9$ at temperature $T/T_{c}\thickapprox0.1$, which shows that the hyperscaling violation really affects the conductivity of the holographic s-wave superconductors. In each panel, the black, blue, green and red lines denote the ground state, first, second and third excited states, respectively. For all cases considered here, we observe that the gap frequency $\omega_{g}$ becomes larger when we increase the value of $\theta$. Obviously, we have larger deviation from $\omega_{g}/T_{c}\approx8$ with the increase of the hyperscaling violation, which indicates that the hyperscaling violation changes the universal relation in the gap frequency $\omega_{g}/T_{c}\approx8$ \cite{GTMM}. In addition, it is interesting to note that there exist $n$ peaks in both imaginary
and real parts of the conductivity for the $n$-th excited state, which is in agreement with the findings in Ref. \cite{WangLLZEPJC}.

\section{Excited states of the p-wave superconductors with hyperscaling violation}

\subsection{Condensates of the vector field}

In this section, we will study the excited states of the p-wave superconductors with the hyperscaling violation via employing the numerical shooting method and analytical Sturm-Liouville approach. In the probe limit, we begin with the Maxwell complex vector field model \cite{CaiPWave-1,CaiPWave-2}
\begin{eqnarray}
S=\int
d^{4}x\sqrt{-g}\left(-\frac{1}{4}F_{\mu\nu}F^{\mu\nu}-\frac{1}{2}\rho_{\mu\nu}^{\dag}\rho^{\mu\nu}-m^2\rho_{\mu}^{\dag}\rho^{\mu}+i
q\gamma\rho_{\mu}\rho_{\nu}^{\dag}F^{\mu\nu}\right),
\end{eqnarray}
where $q$ and $m$ represent the charge and mass of the vector field $\rho_\mu$. The tensor $\rho_{\mu\nu}$ is defined by $\rho_{\mu\nu}=(\nabla_{\mu}-iqA_{\mu})\rho_{\nu}-(\nabla_{\nu}-iqA_{\nu})\rho_{\mu}$ and the strength of $U(1)$ field $A_\mu$ is $F_{\mu\nu}=\nabla_{\mu}A_{\nu}-\nabla_{\nu}A_{\mu}$ \cite{HuangSCPMA}. The parameter $\gamma$ describes the interaction between the vector field $\rho_\mu$ and gauge field $A_\mu$ without external magnetic field. Without loss of generality, we can scale $q=1$ just as in Ref. \cite{CaiPWave-1}. According to the ansatz for the matter fields $\rho_{\upsilon}dx^{\upsilon}=\rho_{x}(u)dx$ and $A_\upsilon dx^{\upsilon}=A_t(u)dt$, we obtain the equations of motion
\begin{eqnarray}
\rho_{x}^{\prime\prime}+\left(\frac{f^\prime}{f}-\frac{1}{u}\right)\rho_{x}^\prime
+\left(\frac{r^{4}_{+}u^{2}A_{t}^2}{f^{2}}-\frac{m^2r_{+}^{\theta}}{u^{2-\theta}f}\right)\rho_{x}=0\,,
\label{NewPWRhox}
\end{eqnarray}
\begin{eqnarray}
A_{t}^{\prime\prime}+\frac{1}{u}A_{t}^\prime-\frac{2r_{+}^{2}\rho_x^2}{f}A_{t}=0,
\label{NewPWPhi}
\end{eqnarray}
where the prime denotes the derivative with respect to $u$. Interestingly, if we set $m^2r_{+}^{\theta}=0$ and rescale the vector field by $\rho_{x}(u)=\psi(u)/\sqrt{2}$ in Eqs. (\ref{NewPWRhox}) and (\ref{NewPWPhi}), we can get the holographic p-wave superconductors with the hyperscaling violation in the $SU(2)$ Yang-Mills model \cite{CaiPWave-5}.

We can obtain the solutions in the superconducting phase by using the boundary conditions for  $\rho_{x}(u)$ and $A_t(u)$. At the horizon $u=1$, the vector field $\rho_{x}(u)$ is required to be regular and the gauge field obeys $A_{t}(1)=0$. And
as $u\rightarrow0$, the asymptotical behaviors are
\begin{eqnarray}\label{RhoxPhiInfinity}
\rho_{x}=\left\{
\begin{array}{rl}
&\rho_{x}^{(2-\Delta)}r_{+}^{2-\Delta}u^{2-\Delta}+\rho_{x}^{(\Delta)}r_{+}^{\Delta}u^{\Delta}\,,~~~~~{\rm
where}\ \Delta=1+\sqrt{1+m^{2}r_{+}^{\theta}}\ {\rm for}\ \theta=0,
\\ \\ &\rho_{x}^{(0)}+\rho_{x}^{(\Delta)}r_{+}^{\Delta}u^{\Delta}\,,~~
\quad\quad\quad\quad\quad\quad\quad  {\rm where}\ \Delta=2\ {\rm
for}\ 0<\theta<2,
\end{array}\right.
\end{eqnarray}
\begin{eqnarray}
A_{t}=\rho+\mu\ln u,
\end{eqnarray}
where $\rho_{x}^{(2-\Delta)}$ (or $\rho_{x}^{(0)}$) and $\rho_{x}^{(\Delta)}$ are interpreted as the source and vacuum expectation value of the vector operator $O_{x}$ in the dual field theory, respectively. Since we are interested in the case where the condensate appears spontaneously, we impose the boundary condition $\rho_{x}^{(2-\Delta)}=0$ (or $\rho_{x}^{(0)}=0$). For concreteness, we will set $m^{2}r_{+}^{\theta}=5/4$ in the following discussion.

From Eqs. (\ref{NewPWRhox}) and (\ref{NewPWPhi}), we obtain the following scaling symmetries
\begin{eqnarray}
&&\rho_{x}\rightarrow\beta^{-1}\rho_{x},\hspace{0.5cm}A_{t}\rightarrow\beta^{-2}A_{t},\nonumber\\
&&\rho_{x}^{(\Delta)}\rightarrow\beta^{-(1+\Delta)}\rho_{x}^{(\Delta)},
\hspace{0.5cm}\mu\rightarrow\beta^{-2}\mu.\hspace{0.5cm}
\label{PWSymmetry}
\end{eqnarray}
where $\beta$ is a positive number. In what follows, we will use them to build the invariant and dimensionless quantities.

\subsubsection{Numerical analysis}

\begin{figure}[ht]
\includegraphics[scale=0.95]{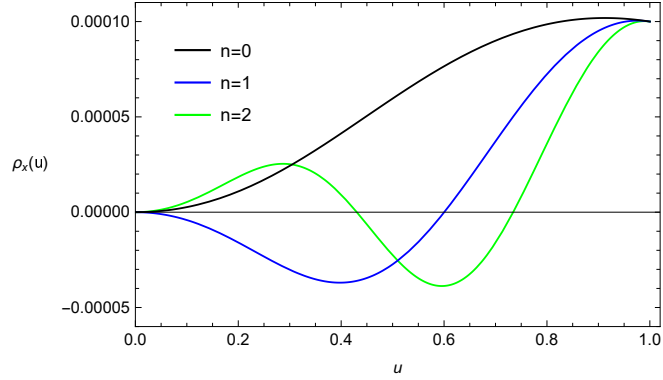}\\ \vspace{0.2cm}
\caption{\label{PNodes}
(Color online.) The distribution of the vector field $\rho_{x}(u)$ as a function of $u$ for the condensate $\langle O_{x}\rangle$ with the hyperscaling violation $\theta=1.0$. The black, blue and green lines denote the ground state, first and second excited states, respectively.}
\end{figure}

In Fig. \ref{PNodes}, we plot the distribution of the vector field $\rho_{x}(u)$  as a function of $u$ for the condensate  $\langle O_{x}\rangle$ with the hyperscaling violation $\theta = 1.0$. Similarly to the case of the scalar field $\psi(u)$, the black, blue and green lines correspond to the ground state $(n=0)$, first excited state $(n=1)$ and second excited state $(n=2)$, respectively. Obviously, there exist exactly $n$ nodes in the $n$-th excited state for both s-wave and p-wave holographic superconductor models.

\begin{figure}[H]
\includegraphics[scale=0.6]{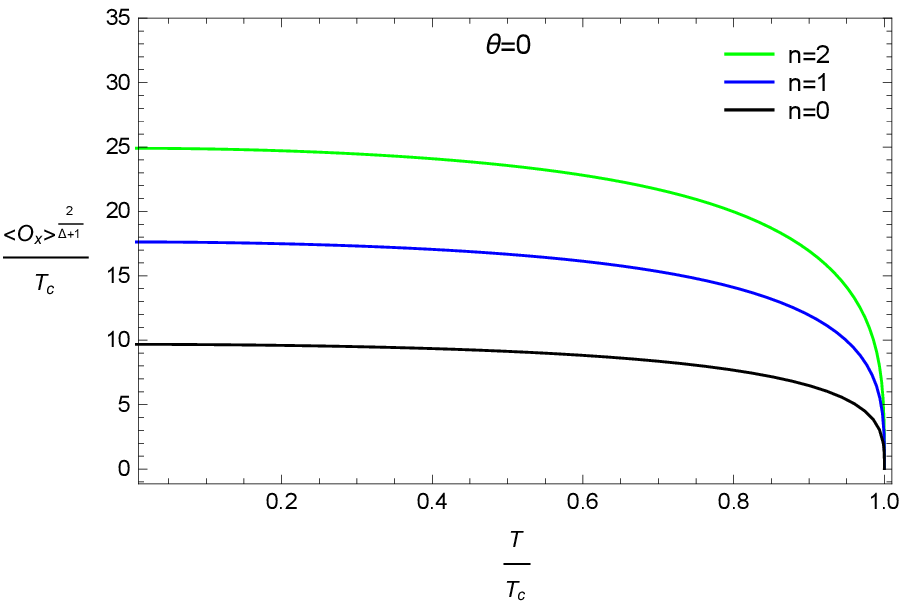}\hspace{0.2cm}%
\includegraphics[scale=0.6]{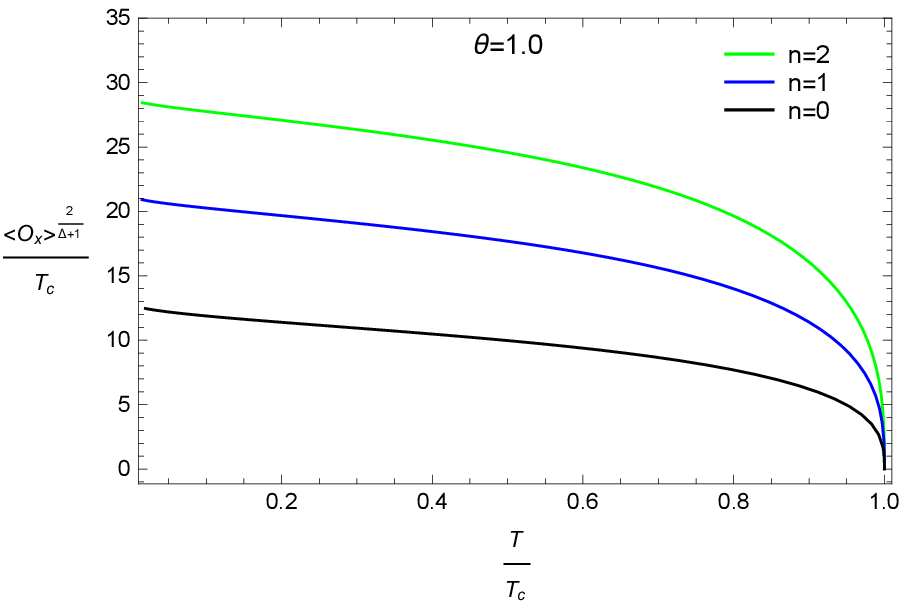}\vspace{0.2cm}%
\includegraphics[scale=0.6]{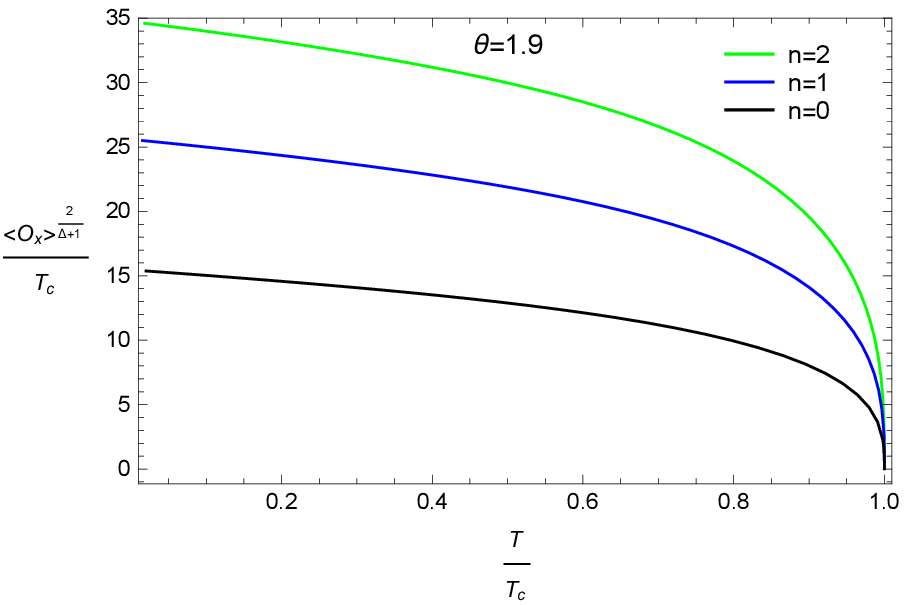}\\ \vspace{0.2cm}
\caption{\label{PwaveCondensate0} (Color online.) The condensate of the vector operator $O_{x}$ with $\theta=0$ (left panel), $\theta=1.0$ (middle panel) and $\theta=1.9$ (right panel) as a function of temperature. In each panel, the black, blue and green lines denote the ground, first and second states, respectively. }
\end{figure}

In Fig. \ref{PwaveCondensate0}, we present the condensate of the vector operator $O_{x}$ as a function of temperature for different $\theta$ in the holographic p-wave superconductor models, which shows that the curve for the operator $O_{x}$ has the similar behavior with the s-wave models for various values of the hyperscaling violation $\theta$. At low temperature, the condensate quickly goes to a fixed value which increases as $n$ increases. By fitting these curves near the critical point, we have
\begin{eqnarray}
\langle O_{0}\rangle\approx
\left\{
\begin{array}{rl}
86.298(T_{c}^{(0)})^{7/4}(1-T/T_{c}^{(0)})^{1/2}, \\ \\
254.955(T_{c}^{(1)})^{7/4}(1-T/T_{c}^{(1)})^{1/2} \ , & \quad  {\rm for}\ \theta=0,\\ \\ 467.700(T_{c}^{(2)})^{7/4}(1-T/T_{c}^{(2)})^{1/2},
\end{array}\right.
\end{eqnarray}
\begin{eqnarray}
\langle O_{1.0}\rangle\approx
\left\{
\begin{array}{rl}
50.270(T_{c}^{(0)})^{3/2}(1-T/T_{c}^{(0)})^{1/2}, \\ \\
124.817(T_{c}^{(1)})^{3/2}(1-T/T_{c}^{(1)})^{1/2} \ , & \quad  {\rm for}\ \theta=1.0,\\ \\ 210.757(T_{c}^{(2)})^{3/2}(1-T/T_{c}^{(2)})^{1/2},
\end{array}\right.
\end{eqnarray}
\begin{eqnarray}
\langle O_{1.9}\rangle\approx
\left\{
\begin{array}{rl}
73.536(T_{c}^{(0)})^{3/2}(1-T/T_{c}^{(0)})^{1/2}, \\ \\
172.469(T_{c}^{(1)})^{3/2}(1-T/T_{c}^{(1)})^{1/2} \ , & \quad  {\rm for}\ \theta=1.9,\\ \\ 283.035(T_{c}^{(2)})^{3/2}(1-T/T_{c}^{(2)})^{1/2},
\end{array}\right.
\end{eqnarray}
where the critical temperatures $T_{c}^{(0)},~T_{c}^{(1)}$, and $T_{c}^{(2)}$ are presented in Table \ref{P-waveNMTcTable} for $n=0,~ n=1$, and $n=2$, respectively. Fitting these curves for all cases considered here, we obtain a square root behavior for small condensate of the operator $O_{x}$
\begin{eqnarray}
\langle O_{x}\rangle\sim(1-T/T_{c}^{(n)})^{1/2}.
\end{eqnarray}
It clearly shows that the holographic p-wave superconducting phase transition with the hyperscaling violation in the excited states is the second order and the critical exponent of the system always takes the mean-field value $1/2$. The hyperscaling violation will not influence the result for all the states.

\begin{table}[ht]
\begin{center}
\caption{\label{P-waveNMTcTable}  The critical temperature $T_{c}$ obtained via the shooting method for the vector operator $O_{x}$ with different values of the hyperscaling violation $\theta$.}
\begin{tabular}{c| c c c c c c c}
         \hline \hline
$\theta$ & 0 & 0.5 & 0.6 & 0.7 & 1.0 & 1.5 & 1.9
        \\
        \hline
$n=0$~~&~~$0.0286863\mu$~~&~~$0.0273161\mu$~~&~~$0.0269773\mu$~~&~~$0.0266180\mu$~~&~~$0.0254286\mu$~~&~~$0.0231662\mu$~~&~~$0.0212113\mu$~
          \\
          \hline
$n=1$~~&~~$0.0152955\mu$~~&~~$0.0144297\mu$~~&~~$0.0142240\mu$~~&~~$0.0140092\mu$~~&~~$0.0133200\mu$~~&~~$0.0120815\mu$~~&~~$0.0110635\mu$~
          \\
          \hline
$n=2$~~&~~$0.0103477\mu$~~&~~$0.0096913\mu$~~&~~$0.0095412\mu$~~&~~$0.0093862\mu$~~&~~$0.0088995\mu$~~&~~$0.0080520\mu$~~&~~$0.0073716\mu$~
          \\
          \hline
$n=3$~~&~~$0.0078013\mu$~~&~~$0.0072717\mu$~~&~~$0.0071535\mu$~~&~~$0.0070325\mu$~~&~~$0.0066571\mu$~~&~~$0.0060151\mu$~~&~~$0.0055059\mu$~
          \\
          \hline
$n=4$~~&~~$0.0062552\mu$~~&~~$0.0058113\mu$~~&~~$0.0057140\mu$~~&~~$0.0056149\mu$~~&~~$0.0053098\mu$~~&~~$0.0047939\mu$~~&~~$0.0043875\mu$~
          \\
          \hline
$n=5$~~&~~$0.0052183\mu$~~&~~$0.0048364\mu$~~&~~$0.0047538\mu$~~&~~$0.0046699\mu$~~&~~$0.0044132\mu$~~&~~$0.0039823\mu$~~&~~$0.0036445\mu$~
          \\
        \hline \hline
\end{tabular}
\end{center}
\end{table}

\begin{figure}[h]
\includegraphics[scale=0.95]{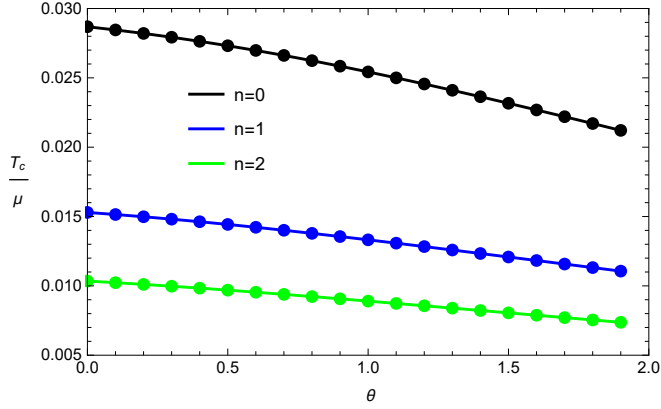}\\ \vspace{0.2cm}
\caption{\label{PTcCondensate}
(Color online.) The critical temperature $T_{c}$ as a function of the hyperscaling violation $\theta$ in the holographic p-wave superconductors. The three lines from top to bottom correspond to the ground (n=0, black), first (n=1, blue) and second (n=2, green) excited states, respectively.}
\end{figure}

In Table \ref{P-waveNMTcTable}, we present the critical temperature for different values of $\theta$ for the p-wave superconducting models. Obviously, the excited state has a lower critical temperature than the ground state for the fixed hyperscaling violation $\theta$. But different from the s-wave superconducting models, when we increase the value of $\theta$, regardless of the ground state and excited states, the critical temperature always decreases, which can be clearly observed from Fig. \ref{PTcCondensate}.

\begin{table}[h]
\caption{\label{Pchepotential} The critical chemical potential $\mu_{c}$ obtained by the shooting method for the vector operator $O_{x}$ from the ground state to the sixth excited state with different values of the hyperscaling violation $\theta$.}
\begin{tabular}{c c c c c c c c}
         \hline
$n$ & 0 & 1 & 2 & 3 & 4 & 5 & 6
        \\
        \hline
~~~~$\theta=0$~~~~&~~~~~$11.096$~~~~~&~~~~~$20.811$~~~~~&~~~~~$30.761$~~~~~&~~~~~$40.802$~~~~
&~~~~~$50.887$~~~~&~~~~~$60.999$~~~~&~~~~~$71.126$~~~~
          \\
~~~~$\theta=1.0$~~~~&~~~~~$9.388$~~~~~&~~~~~$17.923$~~~~~&~~~~~$26.826$~~~~~&~~~~~$35.861$~~~~
&~~~~~$44.961$~~~~&~~~~~$54.095$~~~~&~~~~~$63.252$~~~~
          \\
~~~~$\theta=1.9$~~~~&~~~~~$7.878$~~~~~&~~~~~$15.105$~~~~~&~~~~~$22.670$~~~~~&~~~~~$30.352$~~~~
&~~~~~$38.088$~~~~&~~~~~$45.854$~~~~&~~~~~$53.638$~~~~
          \\
        \hline
\end{tabular}
\end{table}

\begin{figure}[h]
\includegraphics[scale=0.65]{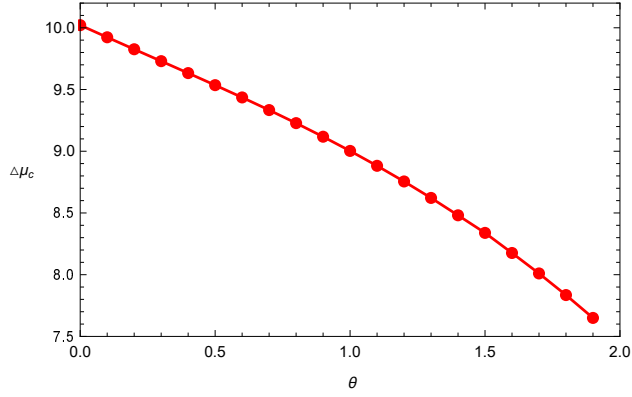}\\ \vspace{0.2cm}
\caption{\label{PChemicalPotential0} (Color online.) The difference of the  dimensionless critical chemical potential between the consecutive states $\Delta\mu_{c}$  as a function of the hyperscaling violation $\theta$ by the shooting method.}
\end{figure}

Also, we want to study the effect of the hyperscaling violation on the difference of the dimensionless critical chemical potential between the consecutive states for the p-wave superconductors. In Table \ref{Pchepotential}, we give the results of $\mu_{c}$ from the ground state to the sixth excited state by using the shooting method for the vector operator $O_{x}$ in the holographic p-wave superconductors. Similarly to the s-wave holographic superconductor models, the excited state has a higher critical chemical potential than the corresponding ground state. Fitting the relation between $\mu_{c}$ and $n$ by using the numerical results, we get
\begin{eqnarray}
\mu_{c}\approx
\left\{
\begin{array}{rl}
10.021n+10.863, \quad {\rm \theta=0}, \\ \\
9.003n+7.486, \quad {\rm \theta=1.0}, \\ \\
7.650n+7.563, \quad {\rm \theta=1.9},
\end{array}\right.
\end{eqnarray}
which indicates that the difference of the dimensionless critical chemical potential depends on the hyperscaling violation. In Fig. \ref{PChemicalPotential0}, we give the difference of the dimensionless critical chemical potential between the consecutive states $\Delta\mu_{c}$ as a function of the hyperscaling violation $\theta$ by using the shooting method, which shows that the difference $\Delta\mu_{c}$ decreases as we amplify $\theta$. It is clear that the hyperscaling violation affects the difference of the dimensionless critical chemical potential between the consecutive states in both the s-wave and p-wave holographic superconductor models.

\subsubsection{Analytical investigation}

Note that the vector field $\rho_{x}=0$ at the critical chemical potential $\mu_{c}$. Thus, as $\mu\rightarrow\mu_{c}$, the equation of motion (\ref{NewPWPhi}) reduces to $A_{t}^{\prime\prime}+\frac{1}{u}A_{t}^\prime=0$, which leads to
\begin{eqnarray}
A_{t}=\lambda r_{+c}^{-2}\ln u \label{PhiSolution},
\end{eqnarray}
with the boundary condition $A_{t}(1)=0$. Working on Eq. (\ref{RhoxPhiInfinity}), we assume that the vector field $\rho_{x}$ takes the form
\begin{eqnarray}
\rho_{x}(u)\sim\langle O_{x}\rangle r_{+}^{\Delta}u^{\Delta}F(u),
\label{BintroduceF}
\end{eqnarray}
where the trial function $F(u)$, with the boundary conditions $F(0)=1$ and $F'(0)=0$, satisfies the following equation of motion
\begin{eqnarray}
(PF^{\prime})^{\prime}+P\left(Q+\lambda^2V\right)F=0,
\end{eqnarray}
with
\begin{eqnarray}
P(u)=u^{2\Delta-1}f,~~Q(u)=\frac{\Delta
}{u}\left(\frac{f'}{f}+\frac{\Delta-2}{u}\right)-\frac{m^2r_{+}^{\theta}}{u^{2-\theta}f},
\end{eqnarray}
where the function $V(u)$ has been defined in Eq. (\ref{BFEoM1}). According to the Sturm-Liouville eigenvalue problem \cite{Gelfand-Fomin}, the eigenvalue $\lambda^{2}$ may be achieved from the extremal values of the following function
\begin{eqnarray}
\lambda^{2}=\left(\mu r_{+c}^{2}\right)^{2}=\frac{\int^{1}_{0}P\left(F'^{2}-QF^{2}\right)du}{\int^{1}_{0}PVF^2du}.
\end{eqnarray}

Considering the boundary conditions $F(0)=1$ and $F'(0)=0$, we choose the trial function as $F(u)=1-\sum_{k=2}^{k=9}a_{k}u^{k}$, just as in the s-wave superconductors with the hyperscaling violation. As an example, we calculate the critical chemical potential $\mu r_{+c}^{2}$ from the ground state to the fifth excited state for the case of the vector field mass $m^{2}r_{+}^{\theta}=5/4$ with the hyperscaling violation $\theta=1.5$, which has been presented in Table \ref{PWaveO1Table1}. According to the Eq. (\ref{lambdaeigenvalue1}), we also obtain the corresponding critical temperature in the Table \ref{PWave} analytically. Obviously, our analytical results agree well with the numerical calculation, i.e., the critical chemical potential $\mu_{c}$ (the critical temperature $T_{c}$) increases (decreases) as the number of nodes $n$ increases for the fixed $\theta$, which indicates that the Sturm-Liouville method is powerful to study the excited states of the p-wave holographic superconducting models with the hyperscaling violation.

\begin{table}[ht]
\begin{center}
\caption{\label{PWaveO1Table1}
The dimensionless critical chemical potential $\mu_{c} r_{+c}^{2}$ for the vector operator $O_{x}$ and the corresponding value of $a_{k}$ for the trial function $F(u)=1-\sum_{k=2}^{k=9}a_{k}u^{k}$. The results of $\mu_{c} r_{+c}^{2}$ are obtained analytically by the Sturm-Liouville method (left column) and numerically by the shooting method (right column) with $\theta=1.5$ from the ground state to the fifth excited state.}
\begin{tabular}{c|c| c c c c c c c c c}
\hline
 $n$&$\mu_{c} r_{+c}^{2}$~&~$a_{2}$~&~$a_{3}$~&~$a_{4}$~&~$a_{5}$~&~$a_{6}$~&~$a_{7}$~&~$a_{8}$~&~$a_{9}$  \\
\hline
$0$&~8.588~~8.588~&~-1.094~&~5.810~&~-7.733~&~7.543~&~-8.737~&~8.158~&~-4.223~&~0.893\\
\hline
$1$&~16.467~~16.467~&~-0.958~&~9.020~&~25.187~&~-99.544~&~105.124~&~-36.331~&~-5.917~&~4.677 \\
\hline
$2$&~24.708~~24.707~&~5.636~&~-67.805~&~544.476~&~-1677.889~&~2511.925~&~-1984.794~&~797.043~&~-127.791\\
\hline
$3$&~33.076~~33.074~&~21.701~&~-316.001~&~2381.473~&~-8292.156~&~15035.304~&~-14851.666~&~7613.893~&~-1591.386 \\
\hline
$4$&~41.789~~41.499~&~-178.178~&~1909.177~&~-7142.562~&~12054.481~&~-8020.209~&~-1761.159~&~4878.861~&~-1739.666 \\
\hline
$5$&~50.964~~49.957~&~-1772.595~&~22587.688~&~-115643.070~&~313237.584~&~-489011.139~&~442649.470~&~-216107.970~&~44061.654 \\
\hline
\end{tabular}
\end{center}
\end{table}

\begin{table}[ht]
\caption{\label{PWave} The critical temperature $T_{c}$ obtained numerically by the shooting method ($T_{cn}$) and analytically by the Strum-Liouville method ($T_{ca}$) for the vector operator $O_{x}$ with $\theta=1.5$ from the ground state to the fifth excited state.}
\begin{tabular}{c c c c c c c c}
         \hline
$n$ & 0 & 1 & 2 & 3 & 4 & 5
        \\
        \hline
~~~~$T_{cn}$~~~~&~~~~$0.0231662\mu$~~~~&~~~~~$0.0120815\mu$~~~~&~~~~$0.0080520\mu$~~~~&~~~~$0.0060151\mu$~~~~&~~~~$0.0047939\mu$~~~~
&~~~~$0.0039823\mu$
          \\
~~~~$T_{ca}$~~~~&~~~~$0.0231662\mu$~~~~&~~~~$0.0120815\mu$~~~~&~~~~$0.0080520\mu$~~~~&~~~~$0.0060147\mu$~~~~
&~~~~$0.0047607\mu$~~~~&~~~~$0.0039036\mu$
          \\
        \hline
\end{tabular}
\end{table}

We will continue to investigate the critical phenomena of the holographic p-wave superconductors with the hyperscaling violation. Since the condensate of the vector field is so small when $T\rightarrow T_{c}$, we will expand $A_{t}(u)$ in terms of small $\langle O_{x}\rangle$ as
\begin{eqnarray}\label{PsavePhiExpandNearTc1}
A_t(u)=\lambda r_{+}^{-2}\ln
u+r_{+}^{2\Delta}\langle O_{x}\rangle^{2}\chi(u)+\cdot\cdot\cdot,
\end{eqnarray}
with the boundary conditions $\chi(1)=0$ and $\chi'(1)=0$. Using Eqs. (\ref{NewPWPhi}), (\ref{BintroduceF}) and (\ref{PsavePhiExpandNearTc1}), we can obtain the equation of motion for $\chi(u)$
\begin{eqnarray}\label{BHChiuEoM1}
(u\chi^\prime)^\prime-\frac{2\lambda u^{2\Delta+1}F^{2}\ln
u}{f}=0.
\end{eqnarray}
Making integration for both sides of Eq. (\ref{BHChiuEoM1}), we have
\begin{eqnarray}
(u\chi^\prime)|_{u\rightarrow
0}=\lambda\mathcal{B}=-\lambda\int^{1}_{0}\frac{2u^{2\Delta+1}F^{2}\ln
u}{f}du.
\end{eqnarray}

From Eqs. (\ref{PhiSolution}) and (\ref{PsavePhiExpandNearTc1}), near $u\rightarrow0$ we arrive at
\begin{eqnarray}
\mu=\lambda
r_{+}^{-2}+r_{+}^{2\Delta}\langle O_{x}\rangle^{2}(u\chi^\prime)|_{u\rightarrow 0},
\end{eqnarray}
which leads to
\begin{eqnarray}\label{OExp1}
\langle O_{x}\rangle=\frac{1}{\sqrt{\mathcal{B}}}\left(\frac{4\pi
T_{c}}{4-\theta}\right)^{\frac{\Delta+1}{2}}\left(1-\frac{T}{T_c}\right)^{\frac{1}{2}},
\end{eqnarray}
which is valid for all cases considered here. To be specific, for the hyperscaling violation $\theta=1.5$, as $T\rightarrow T_{c}$ we get the first three lowest-lying modes
\begin{eqnarray}\label{PWaveO2}
\langle O_{x}\rangle\approx
\left\{
\begin{array}{rl}
48.5(T_{c}^{(0)})^{3/2}(1-T/T_{c}^{(0)})^{1/2}   \ , &  \quad {\rm ground~state}\,, \\ \\
101.0(T_{c}^{(1)})^{3/2}(1-T/T_{c}^{(1)})^{1/2}   \ , &  \quad {\rm 1st~excited~state}\,, \\ \\ 158.1(T_{c}^{(2)})^{3/2}(1-T/T_{c}^{(2)})^{1/2}   \ , &  \quad {\rm 2nd~excited~state}\,,
\end{array}\right.
\end{eqnarray}
with the critical temperatures $T_{c}^{(0)}$, $T_{c}^{(1)}$ and $T_{c}^{(2)}$ given in Table \ref{PWave}, which correspond to the ground, first and second excited states, respectively. Obviously, these results are consistent with the numerical calculation by the shooting method. Therefore, from Eq. (\ref{OExp1}) we can obtain  $\langle{\cal O}_{x}\rangle\sim(1-T/T_{c}^{(n)})^{1/2}$, which analytically confirms that the holographic p-wave superconductor phase transition with the hyperscaling violation in the excited states is of the second order and the critical exponent of the system always takes the mean-field value $1/2$. This behavior is reminiscent of that seen for the holographic s-wave superconductors, so we conclude that the hyperscaling violation will not influence the critical phenomena for the excited states of the holographic s- and p-wave superconductors.

\subsection{Conductivity}

Now we want to know the influence of the hyperscaling violation on the conductivity in the holographic p-wave superconductors. Assuming that perturbed Maxwell field $\delta A_{y}=A_{y}(u)e^{-i\omega t}dy$, we can get the equation of motion for $A_{y}$
\begin{eqnarray}
A_{y}^{\prime\prime}+\left(\frac{f^\prime}{f}-\frac{1}{u}\right)A_{y}^\prime
+\left(\frac{r_{+}^{4}\omega^2u^{2}}{f^2}-\frac{2r_{+}^{2}\rho_{x}^{2}}{f}\right)A_{y}=0,
\label{PWConductivityEquation}
\end{eqnarray}
which can be used to calculate the conductivity. At the horizon, we impose the ingoing wave condition $A_{y}(u)\sim (1-u)^{-i\omega/(4\pi T)}$. And at the boundary $u\rightarrow0$, the general behavior of $A_{y}$ is of the form
\begin{eqnarray}
A_{y}=A_{y}^{(0)}+A_{y}^{(1)}r_{+}^{2}u^{2}.
\end{eqnarray}
Using the AdS/CFT dictionary, we find that the conductivity of the dual superconductor can be expressed as
\begin{eqnarray}
\sigma=\frac{A_{y}^{(1)}}{i\omega A_{y}^{(0)}}\ .
\end{eqnarray}
For different values of the hyperscaling violation, we can obtain the conductivity by solving Eq. (\ref{PWConductivityEquation}) numerically.

\begin{figure}[htbp]
\includegraphics[scale=0.45]{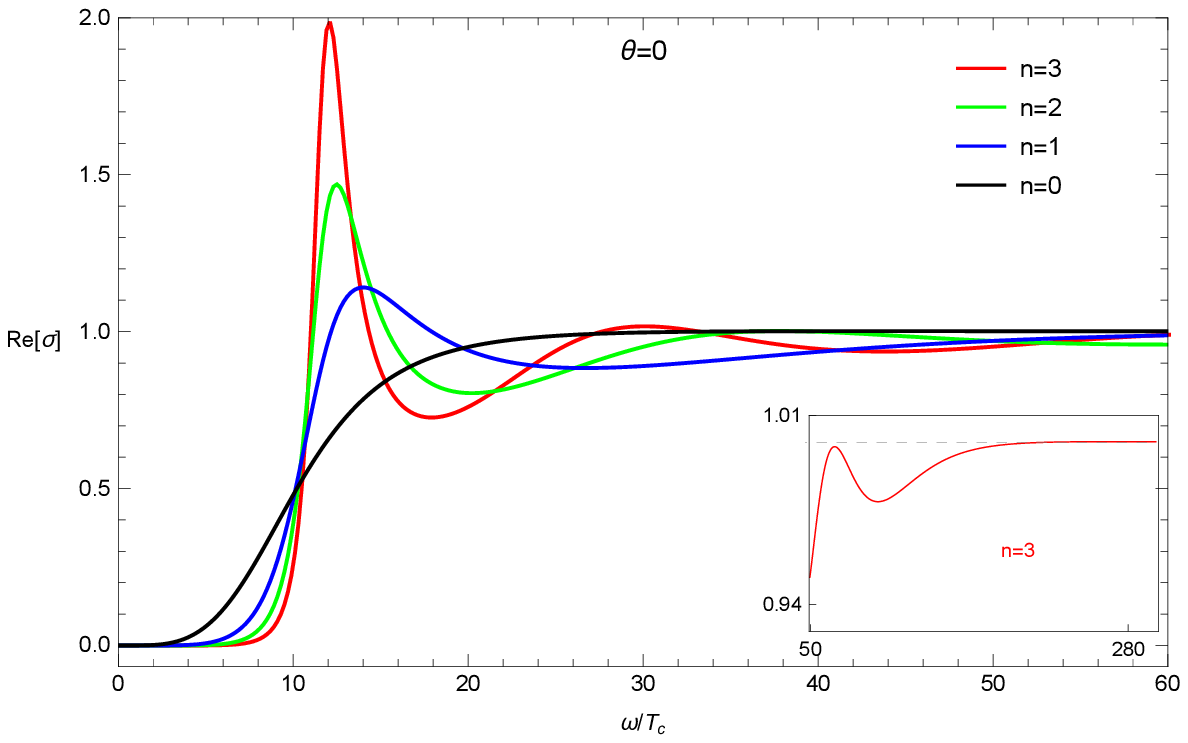}\hspace{0.2cm}%
\includegraphics[scale=0.53]{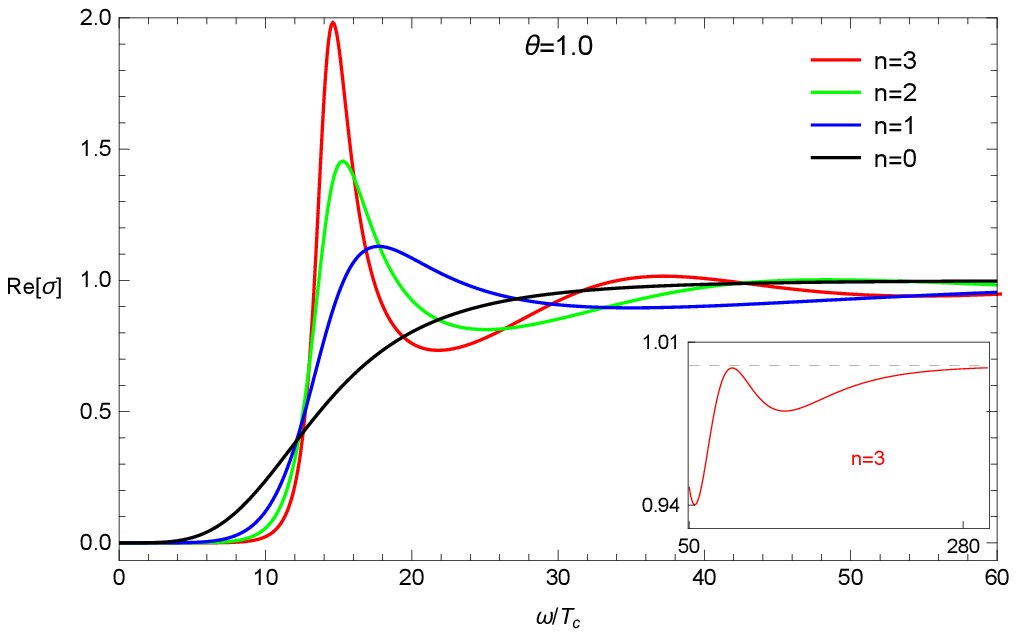}\hspace{0.2cm}%
\includegraphics[scale=0.47]{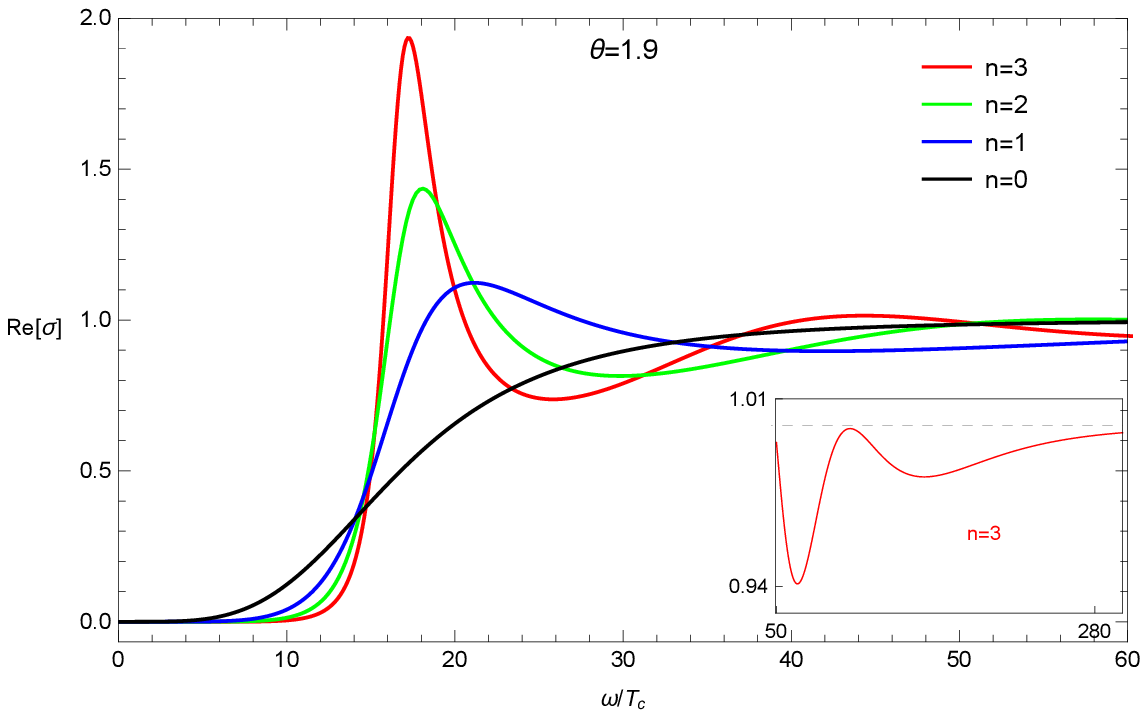}\\ \vspace{0.0cm}
\includegraphics[scale=0.47]{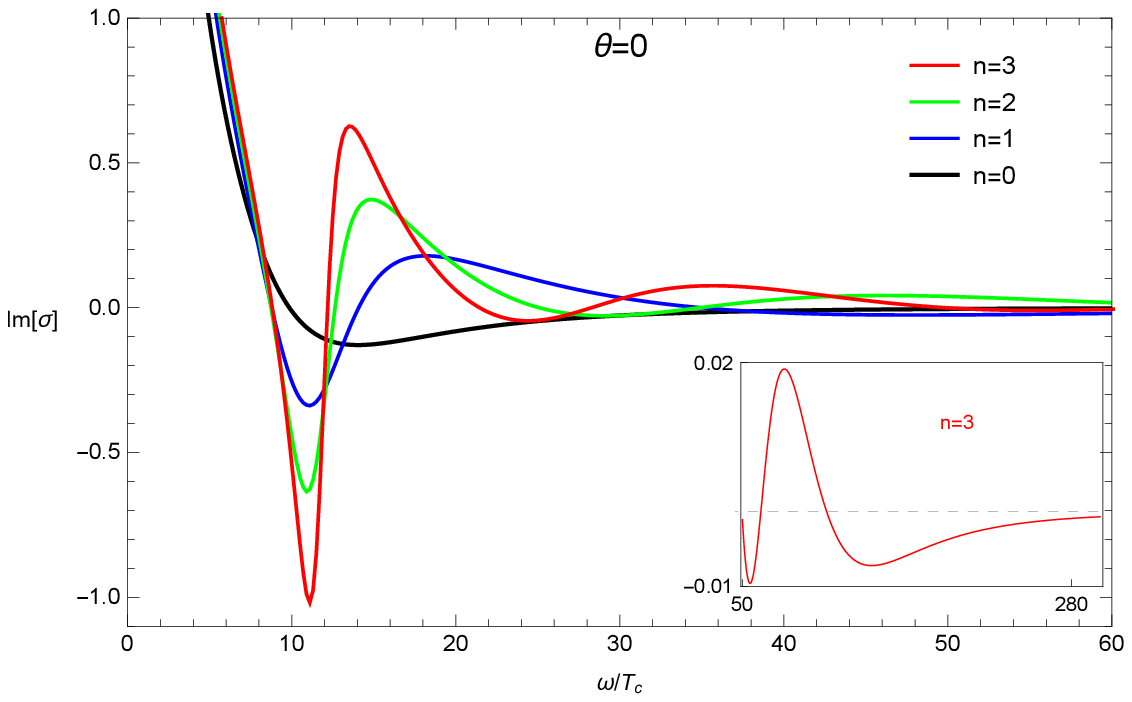}\hspace{0.2cm}%
\includegraphics[scale=0.47]{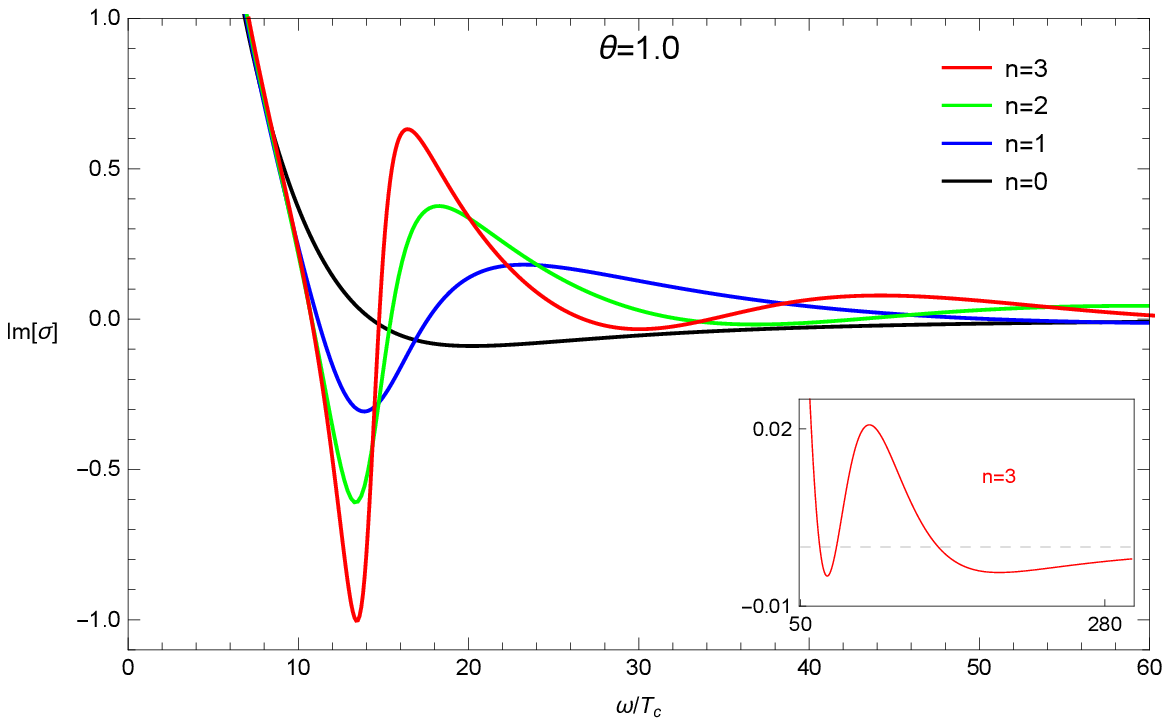}\hspace{0.2cm}%
\includegraphics[scale=0.48]{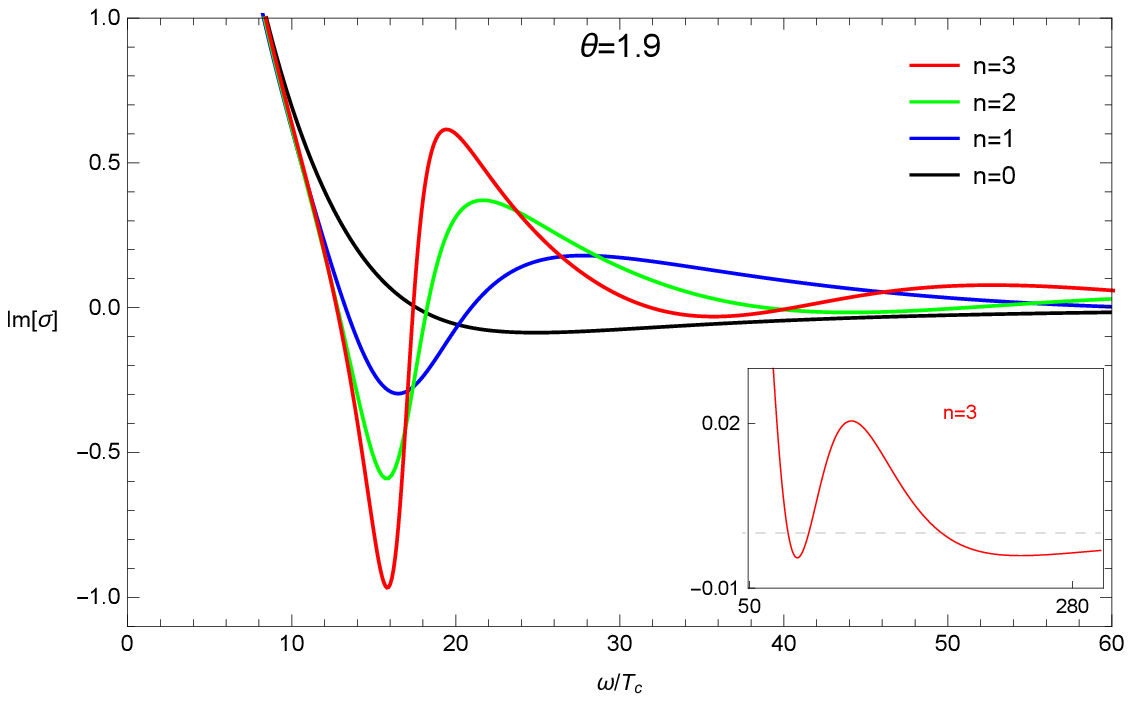}\\ \vspace{0.0cm}
\caption{\label{Conductivity}(Color online.)  The conductivity of the holographic p-wave superconductors for different values of the hyperscaling violation. In each panel, the black, blue, green and red lines denote the ground state, first, second and third excited states, respectively.}
\end{figure}

In Fig. \ref{Conductivity}, for the case of the vector field mass $m^{2}r_{+}^{\theta}=5/4$, we plot the frequency dependent conductivity obtained by solving Eq. (\ref{PWConductivityEquation}) numerically for $\theta=0$, $1.0$ and $1.9$ at temperature $T/T_{c}\approx0.1$. Similarly to the s-wave case, we see that the higher hyperscaling violation results in the larger deviation from the universal value $\omega_{g}/T_{c}\approx8$ \cite{GTMM}. Interestingly, there exist $n$ peaks in both imaginary
and real parts of the conductivity for the $n$-th excited state, which shows that, for the excited states, there exist multiple peaks in both s-wave and p-wave models.

\section{Conclusions}

In the probe limit, we have studied the excited states of holographic superconductors with the hyperscaling violation by the numerical shooting method and analytical Sturm-Liouville method for $d=2,~z=2$, which can help to understand the condensed matter materials with the nonrelativistic symmetry. In the s-wave model, we found that the critical temperature of the ground state decreases first and then increases with the increase of the hyperscaling violation, but the critical temperature of the excited states always decreases. However, in the holographic p-wave superconductor model, we observed that, regardless of the ground state and excited states, the critical temperature always decreases as the hyperscaling violation increases, which is obviously different from the s-wave case. It should be noted that, similar to the relativistic case, the excited state has a lower critical temperature than the corresponding ground state in both s-wave and p-wave models with the hyperscaling violation, which shows that the condensate of the excited states becomes difficult for the operators $O$ (s-wave) and $O_{x}$ (p-wave). For both s-wave and p-wave superconductor models, we found that the critical chemical potential increases linearly with the increasing number of nodes, which means that the excited state has a higher critical chemical potential than the corresponding ground state. However, the difference of the dimensionless critical chemical potential between the consecutive states $\Delta\mu_{c}$ decreases as the hyperscaling violation $\theta$ increases. Moreover, for all the excited states in both s-wave and p-wave models, the holographic superconductor phase transition with the hyperscaling violation is of the second order and the critical exponent of the system always takes the mean-field value $1/2$. The hyperscaling violation will not influence this result. In addition, we investigated the effect of the hyperscaling violation on the conductivity in the excited states and observed that the higher hyperscaling violation results in the larger deviation from the universal value $\omega_{g}/T_{c}\approx8$ in both s-wave and p-wave models. It is interesting to point out that there are the multiple peaks in the conductivity $\sigma$ of the excited states, i.e., there exist $n$ peaks in both imaginary and real parts of the conductivity for the $n$-th excited state. It should be noted that we only considered a particular case with $d=2,~z=2$ in our study. The extension of this work to other cases would be interesting since the combination of the hyperscaling violation exponent $\theta$ and dynamical exponent $z$ can provide richer physics in the holographic superconductor models. We will leave it for further study.

\begin{acknowledgments}

This work was supported by the National Natural Science Foundation of China under Grants Nos. 11705144, 12035005, 11875025, 11775076 and 11690034.

\end{acknowledgments}

\end{document}